\documentclass[paper,notoc,nohyper]{JHEP}
\usepackage{epsfig}
\usepackage{amsbsy} 
\def\S{S_{\epsilon}}

\def\bzero{\beta_0}

\def\Bfin{{\rm Box^6} (u, t) }
\def\Bubl{{\rm Bub}(s)}
\def\aa{{\bom A}(\ep,s,t,u)}
\def\aatwo{{\bom A}(2\ep,s,t,u)}
\def\bb{{\bom B}(\ep,s,t,u)}
\def\bbtwo{{\bom B}(2\ep,s,t,u)}
\def\cc{{\bom C}(\ep,s,t,u)}

\def\dd{{\bom D}(\ep,s,t,u)}
\def\ddb{{\bom D}(\ep,s,u,t)}

\def\cone{\frac{\beta_0}{\epsilon}}
\def\Tone{{\cal T}_1}
\def\Ttwo{{\cal T}_2}
\def\Lone{{\cal L}_1}
\def\Ltwo{{\cal L}_2}
\def\Lthree{{\cal L}_3}
\def\Htwo{{\cal H}_2}
\def\R{{\cal R}}

\def\lnx{X}
\def\lny{Y}
\def\Ls{S}
\def\Poles{{\cal P}oles}
\def\Finite{{\cal F}inite}
\def\Libx{{\rm Li}_2(x)}
\def\Liby{{\rm Li}_2(y)}
\def\Licx{{\rm Li}_3(x)}
\def\Licy{{\rm Li}_3(y)}
\def\Lidx{{\rm Li}_4(x)}
\def\Lidy{{\rm Li}_4(y)}
\def\Lidz{{\rm Li}_4\Biggl(\frac{x-1}{x}\Biggr)}

\def\CA{C_A}
\def\CF{C_F}

\def\NF{N_F}

\def\C{{\cal C}}
\renewcommand\O[1]{{\cal O}\left(#1\right)}
\def\as{\ensuremath{\alpha_{s}}}
\def\a0{\alpha_0}

\def\Re{\mathop{\rm Re}}

\def\qb{\bar q}

\def\beq{\begin{equation}}
\def\eeq{\end{equation}}

\def\beqn{\begin{eqnarray}}
\def\eeqn{\end{eqnarray}}

\def\lq{\left[}
\def\rq{\right]}

\def\({\left(}
\def\){\right)}

\def\ket#1{|{#1}\rangle}
\def\bra#1{\langle{#1}|}
\def\braket#1#2{\langle #1 |#2 \rangle}
\def\cm{{\cal M}}

\def\MSbar{$\overline{{\rm MS}}$}

\def\bom#1{{\mbox{\boldmath $#1$}}}

\def\fs{\(-\frac{\mu^2}{s}\)^\ep }
\def\ft{\(-\frac{\mu^2}{t}\)^\ep }
\def\fu{\(-\frac{\mu^2}{u}\)^\ep }

\def \ep{\epsilon}

\def\absq#1{ {\left | #1 \right |}^2}
\def\boxs#1{{\rm Box}^6 \left( #1 \right)}
\def\dag{\dagger}
\def\bnot{\beta_0}
\def\bubs{{\rm Bub}\left( s\right)}
\def\bubt{{\rm Bub}\left( t\right)}
\def\bubu{{\rm Bub}\left( u\right)}

\def\bst{\boxs{s, t}}
\def\bsu{\boxs{s, u}}
\def\btu{\boxs{t, u}}
\def\fbub#1{ {\cal F_E} \left(#1\right)}
\def\MAA{{\cal F_D}}
\def\MAB{{\cal F_C}}
\def\MCA{{\cal F_B}}
\def\MCC{{\cal F_A}}
\def\PC{{\cal P_C}}
\def\PA{{\cal P_A}}
\def\RA{{\cal R_A}}
\def\PSQUARE{{\cal P_S}(s,t,u)}
\def\PLOOP{{\cal P_L}(s,t,u)}
\def\FINITE{{\cal F_T}(s,t,u)}
\def\PCC{\PC_{,C}}
\def\PAC{\PA_{,C}}

\def\AA{{\bom A}}
\def\BB{{\bom B}}

\def\Ls{L_s}
\def\Lu{L_u}
\def\Lx{L_x}
\def\Ly{L_y}
\def\Lx{X}
\def\Ly{Y}
\def\Ls{S}
\def\Lu{U}
\def\Libx{{\rm Li}_2(x)}
\def\Liby{{\rm Li}_2(y)}
\def\Licx{{\rm Li}_3(x)}
\def\Licy{{\rm Li}_3(y)}

\def\Lidx{{\rm Li}_4(x)}
\def\Lidy{{\rm Li}_4(y)}
\def\Lidz{{\rm Li}_4(z)}

\def\sstt{\frac{t}{u}}

\def\utss{ }

\def\tttu{\frac{t^2}{s^2}}
\def\tfiveou{\frac{t^4}{u^2s^2}}

\def\uutps{\left[\frac{t^2+s^2}{st}\right]}
\def\uutms{\left[\frac{t^2-s^2}{st}\right]}
\def\stuu{}

\def\tsmst{\left[\frac{t^2-s^2}{u^2}\right]}
\def\tspst{\left[\frac{t^2+s^2}{u^2}\right]}

\def\tfiveos{\frac{t^4}{s^2u^2}}
\def\sfiveot{\frac{s^4}{t^2u^2}}

\title{\boldmath 
Two-loop 
QCD corrections to massless quark-gluon scattering\footnote{Work supported 
in part by the UK Particle Physics and
Astronomy Research Council and by the EU Fourth Framework Programme
`Training and Mobility of Researchers', Network `Quantum Chromodynamics
and the Deep Structure of Elementary Particles',
contract FMRX-CT98-0194 (DG 12 - MIHT).
C.A. acknowledges
the financial support of the Greek government and
M.E.T. acknowledges financial support
from CONACyT and the CVCP. We thank
the British Council and German Academic Exchange Service for support
under ARC project 1050.}
}
\author{
C.~Anastasiou$^a$,
E.~W.~N.~Glover$^a$,
C.~Oleari$^b$ and M.~E.~Tejeda-Yeomans$^a$\\
$^a$Department of Physics, 
University of Durham, 
Durham DH1 3LE, 
England\\[1mm]
$^b$Department of Physics, 
University of Wisconsin,
1150 University Avenue\\
Madison WI 53706, 
U.S.A.\\[1mm]
E-mail: \email{Ch.Anastasiou@durham.ac.uk}, \email{E.W.N.Glover@durham.ac.uk},
\email{Oleari@pheno.physics.wisc.edu}, 
\email{M.E.Tejeda-Yeomans@durham.ac.uk}}
\abstract{
We present the $\O{\as^4}$ virtual QCD corrections to  the scattering process
of massless quark $q \bar q \to gg$
due to the interference of tree and two-loop amplitudes and
to the self-interference of one-loop amplitudes. We work in conventional
dimensional regularisation and our results are renormalised in the \MSbar\
scheme.  The structure of the infrared divergences agrees with that predicted
by Catani while expressions for the finite remainder are given for the $q \bar
q \to gg$ and the $q g \to q g$ ($g\bar q \to g \bar q$) scattering processes
in terms of logarithms and  polylogarithms that are real in the physical
region.   These results, together with those previously obtained for
quark-quark scattering, are important ingredients in the
next-to-next-to-leading order contribution to inclusive jet production at
hadron colliders.  }

\preprint{{DCTP/01/04}, {IPPP/01/02}, {MADPH-00-1210}, {hep-ph/yymmnnn}}

\begin{document}

\section{Introduction}
\label{sec:intro}

In hadronic collisions, jet cross sections are computed as a convolution of 
hard partonic cross sections with parton-distribution functions,  followed by
the fragmentation of the final-state partons into hadrons. The theoretical
prediction can be improved by including higher order corrections which have the
effect of reducing the unphysical renormalisation- and factorisation-scale
dependences and by improving the matching of the parton-level theoretical jet
algorithm with the hadron-level experimental jet algorithm.  At present jet
production is described at next-to-leading order (NLO) and  several numerical
programs are available~\cite{EKS,jetrad} which have been extensively used
to compare with data from the TEVATRON and CERN S$p\bar p$S.

Improving the theoretical prediction to next-to-next-to-leading order (NNLO)
requires several ingredients.  First, the  parton-density functions are 
needed to NNLO accuracy which in turn  requires
knowledge of the three-loop splitting functions. At present, the even moments
of the splitting functions are known for the flavour singlet and non-singlet
structure functions $F_2$ and $F_L$ up to $N=12$  while the odd moments up to
$N=13$ are known for $F_3$ \cite{moms1,moms2}. The numerically small $\NF^2$
non-singlet contribution is also known~\cite{Gra1}. These moments are
sufficient to parameterise  the splitting functions in
$x$-space~\cite{NV,NVplb} and NNLO global analyses~\cite{MRS} are starting to
appear. Second, the hard scattering matrix elements  can be computed at NNLO.
In the high energy limit, the quarks can be assumed to be massless and at
this order, there
are contributions from $2 \to 4$ tree-level diagrams~\cite{6g,4g2q,2g4q,6q},
from $2\to 3$ one-loop-level  diagrams~\cite{5g,3g2q,1g4q}  and two-loop $2\to
2$ diagrams. 

The evaluation of the two-loop diagrams has been a challenge for the past few
years due to the presence of two-loop planar and crossed boxes.  In the
massless parton limit and in dimensional regularisation, analytic expressions  
for these basic scalar integrals  have now been provided by
Smirnov~\cite{planarA} and Tausk~\cite{nonplanarA}  as series  in
$\ep=(4-D)/2$, where $D$ is the space-time dimension,
together with constructive procedures for reducing tensor integral to a basis
set of known scalar (master) integrals~\cite{planarB,nonplanarB}.  This makes
the calculation of the two-loop amplitudes for $2 \to 2$ QCD scattering
processes possible. Bern, Dixon and Kosower~\cite{bdk} were the first to
address such scattering processes and provided analytic expressions for the
maximal-helicity-violating two-loop amplitude for $gg \to gg$. Subsequently,
Bern, Dixon and Ghinculov~\cite{BDG}  completed the two-loop calculation of
physical $2 \to 2$ scattering amplitudes for the QED  processes $e^+e^- \to
\mu^+\mu^-$ and $e^+e^- \to e^-e^+$.  More recently, we have provided
expressions relevant  for unlike- and like-quark scattering in the massless
limit in Refs.~\cite{qqQQ,qqqq} respectively.  The corresponding
matrix elements for the self-interference of one-loop $2 \to 2$ quark
scattering processes are given in~\cite{1loopsquare}.

In this paper, we address the $\O{\as^4}$ one- and two-loop corrections to the
QCD process
\begin{equation} 
q + \bar q \to  g  + g,
\label{eq:qqgg}
\end{equation}
together with the time-reversed and crossed processes
\begin{eqnarray}
q + g&\to & q + g,\\ 
g + \bar{q} &\to & g +\bar{q},\\
g + g &\to& q + \bar{q}. \label{eq:qgqg} 
\end{eqnarray} 
As is in Refs.~\cite{qqQQ,qqqq,1loopsquare},  we use the \MSbar\
renormalisation 
scheme to remove the ultraviolet singularities and conventional dimensional
regularisation, where all external particles are treated in $D$ dimensions.  We
provide expressions for both the interference of tree-level  and two-loop
graphs as 
well as the self-interference of one-loop amplitudes. In each case, we find
that the infrared pole structure  agrees with that obtained using Catani's
general factorisation formulae~\cite{catani}.   The finite remainders are the
main new results presented in this paper and we give explicit analytic
expressions valid for each of the processes  of
Eqs.~(\ref{eq:qqgg})--(\ref{eq:qgqg}) in 
terms of logarithms and polylogarithms  that are real in the physical domain.  
For simplicity, we decompose our results according to the powers of the number
of colours $N$ and the number of light-quark flavours $\NF$ .

Our paper is organised as follows.  We first establish our notation in
Sec.~\ref{sec:notation}. Analytic expressions for the interference of the
two-loop and tree-level amplitudes are given in Sec.~\ref{sec:two}, while
formulae describing the self-interference of one-loop graphs are given in
Sec.~\ref{sec:one}.  In Sec.~\ref{subsec:polestwo} we adopt the notation
used in Ref.~\cite{catani}, to isolate the infrared singularity structure of
the two-loop amplitudes in the \MSbar\ scheme and we demonstrate that the
anticipated singularity structure agrees with our explicit calculation. The
finite $\O{\ep^0}$ remainder of the two-loop graphs is one of the main
results of our paper and expressions appropriate for the $q \bar q \to g g $,
$qg \to qg$ ($g\bar q \to g \bar q$) and $gg \to q \bar q$ scattering
processes are given in Sec.~\ref{subsec:finitetwo} in terms of logarithms
and polylogarithms that have no imaginary parts.  Expressions for the
self-interference of one-loop graphs are given in Sec.~\ref{sec:one} in
terms of the one-loop bubble integral in $D=4-2\ep$ and the one-loop box
integral in $D=6-2\ep$.  Analytic formulae connecting these integrals in the
various kinematic regions are given in Appendix~\ref{app:master_int}.  As for
the two-loop contributions, the singularity structure agrees with that
obtained using Catani's formalism.  Finally we conclude with a brief summary
of the results in Sec.~\ref{sec:conc}.

\section{Notation} 
\label{sec:notation}
For calculational purposes, the process we consider is
\begin{equation}
\label{eq:proc}
q (p_1) + \bar q (p_2)  + g (p_3) + g(p_4) \to 0,
\end{equation}
where the particles are all incoming and carry light-like momenta, 
satisfying 
$$
p_1^\mu+p_2^\mu+p_3^\mu+p_4^\mu = 0, \qquad p_i^2=0.
$$
The associated Mandelstam variables are given by
\begin{equation}
s = (p_1+p_2)^2, \qquad t = (p_2+p_3)^2, \qquad u = (p_1+p_3)^2, 
\qquad s+t+u = 0.
\end{equation}
We work in conventional dimensional regularisation  treating all external 
quark and gluon states in $D$ dimensions  and renormalise the ultraviolet 
divergences in the \MSbar\ scheme. The bare coupling $\a0$ is related to 
the running coupling $\as \equiv \alpha_s(\mu^2)$  at renormalisation 
scale $\mu$, by
\beq
\label{eq:alpha}
\a0 \,  \S = \as \,  \lq 1 - \frac{\beta_0}{\ep}  
\, \left(\frac{\as}{2\pi}\right) + \( \frac{\beta_0^2}{\ep^2} - \frac{\beta_1}{2\ep} \)  \, 
\left(\frac{\as}{2\pi}\right)^2
+\O{\as^3} \rq,
\eeq
where
\beq
\S = (4 \pi)^\ep e^{-\ep \gamma},  \quad\quad \gamma=0.5772\ldots=
{\rm Euler\ constant}
\eeq
is the typical phase-space volume factor in $D=4-2\ep$ dimensions.
The first two coefficients of the QCD beta function, 
$\beta_0$ and $\beta_1$, for $N_F$ (massless) quark flavours, are
\beq
\label{betas}
\beta_0 = \frac{11 \CA - 4 T_R \NF}{6} \;\;, \;\; \;\;\;\;
\beta_1 = \frac{17 \CA^2 - 10 \CA T_R \NF - 6 \CF T_R \NF}{6} \;\;,
\eeq
where $N$ is the number of colours and 
\beq
\CF = \frac{N^2-1}{2N}, \qquad \CA = N, \qquad T_R = \frac{1}{2}.
\eeq
The renormalised four point amplitude in the \MSbar\  scheme is thus
\beqn
\ket{\cm}&=& 4\pi \as \Biggl [  \ket{\cm^{(0)}}  
+ \left(\frac{\as}{2\pi}\right)  \ket{\cm^{(1)}} 
+ \left(\frac{\as}{2\pi}\right)^2\,
 \ket{ \cm^{(2)}}  + \O{\as^3} \Biggr ],\nonumber \\
\eeqn
where the $\ket{\cm^{(i)}}$ represents the colour-space vector describing the
$i$-loop amplitude. The dependence on both renormalisation scale $\mu$ and
renormalisation scheme is implicit.

We denote the squared amplitude summed over spins and colours by
\beq
\braket{\cm}{\cm} = \sum |{\cal M}({q + \bar{q} \to  g + g })|^2
= \C(s,t,u).
\eeq
which is symmetric under the exchange of $t$ and $u$.

The squared matrix elements for the crossed processes are 
obtained by exchanging the Mandelstam variables and 
introducing a minus sign for each quark change 
between initial and final states
\beqn
\sum |{\cal M}({g + g \to  q + \bar q })|^2  
&=& \C(s,t,u),\\
\sum |{\cal M}({q + g \to  q + g })|^2  
&=& - \C(u,t,s),\\
\sum |{\cal M}({g + \bar q \to  g + \bar q })|^2  
&=& - \C(u,t,s).
\eeqn
The function $\C$ can be expanded perturbatively to yield
\beqn
\C(s,t,u) &=& 16\pi^2\as^2 \left[
 \C^4(s,t,u)+\left(\frac{\as}{2\pi}\right) \C^6(s,t,u)
 +\left(\frac{\as}{2\pi}\right)^2 \C^8(s,t,u) +
\O{\as^{3}}\right],\nonumber \\  
\eeqn
where
\beqn
\C^4(s,t,u) &=& \braket{\cm^{(0)}}{\cm^{(0)}} =
\nonumber\\
&=& 2\,\frac{N^2-1}{N}(1-\ep)\(\frac{N^2-1}{ut} - 2\frac{N^2}{s^2}\)
\(t^2+u^2-\ep s^2 \),\\
\C^6(s,t,u) &=& \left(
\braket{\cm^{(0)}}{\cm^{(1)}}+\braket{\cm^{(1)}}{\cm^{(0)}}\right),\\
\C^8(s,t,u) &=& \left( \braket{\cm^{(1)}}{\cm^{(1)}} +
\braket{\cm^{(0)}}{\cm^{(2)}}+\braket{\cm^{(2)}}{\cm^{(0)}}\right). 
\eeqn
Expressions for $\C^6$ are given in Ref.~\cite{ES} using dimensional
regularisation to isolate the infrared and ultraviolet singularities.  

In the following sections, we present expressions for the infrared singular 
and finite contributions to $\C^8$ and the crossed processes. 
For convenience, 
we divide $\C^8(s,t,u)$ into two pieces
\begin{itemize}
\item[-] the pure two-loop contributions 
\begin{equation}
\C^{8\, (2 \times 0)}(s,t,u) = \braket{\cm^{(0)}}{\cm^{(2)}}+\braket{\cm^{(2)}}{\cm^{(0)}},
\end{equation} 
described in Sec.~\ref{sec:two} and
\item[-] the self-interference of the one-loop amplitude
\begin{equation}
\label{eq:self_interf}
\C^{8\, (1 \times 1)}(s,t,u) = \braket{\cm^{(1)}}{\cm^{(1)}},
\end{equation} 
described in Sec.~\ref{sec:one}.
\end{itemize}

As in Refs.~\cite{qqQQ,qqqq}, we use {\tt QGRAF}~\cite{QGRAF} to  produce the
two-loop Feynman diagrams to construct $\ket{\cm^{(2)}}$. 
We then project by 
$\bra{\cm^{(0)}}$ and perform the summation over colours and
spins. It should be noted that when summing over the gluon polarisations, 
we ensure
that the polarisations states are transversal (i.e. physical) by using 
an axial
gauge
\begin{equation}
\sum_{{\rm spins}} \ep_{i}^{\mu}\ep_{i}^{\nu\ast} = 
-  g^{\mu \nu} + \frac{n_{i}^{\mu}p_{i}^{\nu} 
+ n_{i}^{\nu}p_{i}^{\mu}}{n_{i} \cdot p_{i}} 
\end{equation}
where $p_{i}$ is the momentum of gluon $i$ and 
$n_{i}$ is an arbitrary light-like 4-vector. 
For simplicity, we choose $n_3^{\mu} = p_4^{\mu}$ and $n_4^{\mu} = p_3^{\mu}$.
Finally, the  trace over the Dirac matrices is carried  out in $D$
dimensions using conventional dimensional regularisation. It is then
straightforward to identify the scalar and tensor integrals present  and
replace them with combinations of the basis set  of master integrals using
the  tensor reduction of two-loop integrals described in
\cite{planarB,nonplanarB,AGO3}, based on integration-by-parts~\cite{IBP} and 
Lorentz invariance~\cite{diffeq} identities.   The final result is  a
combination of master integrals in $D=4-2\epsilon$ for which the 
expansions around $\ep = 0$ are given
in~\cite{planarA,nonplanarA,planarB,nonplanarB,AGO3,AGO2,xtri,bastei3,bastei2}.   

\section{Two-loop contribution}
\label{sec:two}
We further decompose the two-loop contributions as a sum of two terms
\beq
\C^{8 \, (2\times 0)}(s,t,u)
 = \Poles(s,t,u)+\Finite(s,t,u).
\eeq 
$\Poles$ contains infrared singularities that will be  analytically
canceled by the infrared singularities occurring in radiative processes of the
same order (ultraviolet divergences are removed by renormalisation).
$\Finite$ is the remainder which is finite as $\ep \to 0$.

\subsection{Infrared Pole Structure}
\label{subsec:polestwo}
Following the procedure outlined in Ref.~\cite{catani}, we can write the
infrared pole structure, renormalised in the \MSbar\ scheme as
\begin{eqnarray}
\label{eq:poles}
\lefteqn{\Poles(s,t,u) = \Re \Bigg\{} \nonumber \\
 &&\Biggl[
\frac{V^2}{4N}\left(-\aa^2-\bb\dd+2\R\aatwo-4\cone\aa
\right)
\nonumber \\
&&\hspace{1cm}+\frac{V}{4} \Big(-\aa\bb-\bb\cc+2\R\bbtwo
\nonumber \\
&&\hspace{1cm}
-4\cone\bb\Big)
+\frac{V}{4N} \bb\ddb 
\Biggr]\Tone(s,t,u)
\nonumber \\
&&+\Biggl[
\frac{V}{4N}
\left(\aa^2+\bb\dd+4\cone\aa -2\R\aatwo 
\right)
\nonumber \\
&&\hspace{1cm}+\frac{V}{4}
\Big(-\aa\bb-\bb\cc-4\cone\bb\nonumber \\
&&\hspace{1cm}
+2\R\bbtwo \Big)
-\frac{V^2}{4N} \bb\ddb
\Biggr]
\,{\Ttwo}(s,t,u)
\nonumber \\
&&+\left[
\frac{V^2}{4N}\aa + \frac{V}{4} \bb
\right]{\Lone}(s,t,u)
\nonumber \\
&&+\left[
-\frac{V}{4N}\aa + \frac{V}{4} \bb
\right]{\Ltwo}(s,t,u)\nonumber \\
&&+
\left[
\frac{V^2}{4N} \dd + \frac{V}{4} \cc - \frac{V}{4N} \ddb
\right]\,{\Lthree}(s,t,u)  + \Htwo \nonumber \\
&& + \, ( u \leftrightarrow t) \Bigg\} ,
\end{eqnarray}
where
\begin{eqnarray}
\label{eq:Adef}
\aa &=&{ }-{}\Biggl\{{}\Biggl({1\over \ep^2}+{3\over 2  \ep}\Biggr){}\,
{}\Biggl[{N}\,{\fu}+\CF \,{\fs}\Biggr]{}\nonumber\\
&&\phantom{{ }-{}\Biggl\{}
+{N}\,{}\Biggl ({\bzero\over 2N  \ep}-{3\over 4\ep}\Biggr
){}\,{}\Biggl[{\fs}+{\fu}\Biggr]{}\Biggr\} \,
\frac{e^{\ep\gamma}}{\Gamma(1-\ep)}{}\\
\label{eq:Bdef}
\bb &=&{ }{}\Biggl ({1\over \ep^2}+{3\over 4  \ep}+\,{\bzero\over 2 N
\ep}\Biggr 
)\,{}\Biggl[ {\ft} -{\fs}\Biggr]{}\,\frac{e^{\ep\gamma}}{\Gamma(1-\ep)}{}\\
\label{eq:Cdef}
\cc &=&{ }-{}\Biggl [\Biggl ({1\over \ep^2}+{3\over 2 \ep}\Biggr )\CF
+{}\Biggl ({1\over \ep^2}+{\bzero\over N \ep}\Biggr){}
\,{N}\Biggr ]{\fs}\frac{e^{\ep\gamma}}{\Gamma(1-\ep)}{}\\ 
\label{eq:Ddef}
\dd &=&{ }{}\Biggl ({1\over \ep^2}+{3\over 4\ep}+\,{\bzero\over 2 N \ep}\Biggr
)\,{}\Biggl[{\ft}-{\fu}\Biggr ]{}\,\frac{e^{\ep\gamma}}{\Gamma(1-\ep)}{}
\end{eqnarray}
and
\beqn
V &=& N^2 -1,\\
\R &=& e^{-\ep \gamma } \frac{ \Gamma(1-2\ep)}{\Gamma(1-\ep)} 
\left(\frac{\beta_0}{\epsilon} + K\right),\\
K &=& \left( \frac{67}{18} - \frac{\pi^2}{6} \right) \CA - \frac{10}{9} T_R
\NF.
\eeqn 
The tree-type structures, $\Tone$ and $\Ttwo$, are given by
\begin{eqnarray}
\label{eq:Tonedef}
\Tone(s,t,u)&=& \frac{t}{u} \Ttwo(s,t,u),\\
\label{eq:Ttwodef}
\Ttwo(s,t,u)&=& 8 (1-\ep)\frac{\(t^2+u^2-\ep s^2\)}{s^2},
\end{eqnarray}
while the interference of tree with one-loop structures are represented by
$\Lone$, $\Ltwo$ and $\Lthree$ 
\begin{eqnarray}
\label{eq:Lonedef}
\Lone(s,t,u)&=& \frac{N^2+1}{N} f_2(s,t,u) - \frac{t}{N u} f_1(s,u,t) +
f_3(s,t,u)  \nonumber \\ 
&& - 6 \beta_0 \frac{(1-\ep)}{3-2\ep}~ \Tone(s,t,u)~ {\rm Bub}(s)  \nonumber \\
&& - 2 \ep (1-2\ep) \left[ \frac{N }{\ep^2}{\rm Bub}(u) + \frac{N}{2} \left(
\frac{1}{\ep^2} - \frac{2 \beta_0}{N \ep} + \frac{3}{2\ep} \right){\rm
Bub}(s) \right .  \nonumber \\ 
&& \left.- \frac{1}{2N} \left( \frac{1}{\ep^2} + \frac{3}{2 \ep} \right){\rm
Bub}(s) \right] \Tone(s,t,u)  \\  
\label{eq:Ltwodef}
\Ltwo(s,t,u)&=& \frac{t}{u} \Lone(s,u,t) \\
\label{eq:Lthreedef}
\Lthree(s,t,u)&=& 16 (1-2\ep) \frac{1}{u}
\left[ t^2+u^2+\(ut-2t^2-2u^2\)\ep+\(t^2+u^2+3ut\)\ep^2 \right]
{\rm Box}^6(t,u) \nonumber \\ 
&& +2 \(\frac{1}{\ep} -2\) \lq{\rm Bub}(t)-{\rm Bub}(s)\rq \Tone(s,t,u) +
f_1(s,t,u) 
\nonumber \\ 
&& +2\(\frac{1}{\ep} -2\)  \lq{\rm Bub}(u)-{\rm Bub}(s)\rq \Ttwo(s,t,u) +
\frac{t}{u}f_1(s,u,t),
\end{eqnarray}
where the infrared-finite functions $f_1,f_2$ and $f_3$ are 
\begin{eqnarray}
f_1(s,t,u)&=& 8(1-2\ep)\frac{1}{s^2}
\Big[u\(2u^2+5t^2+3tu\)+\ep\(3t^3-4u^3-3tu^2\)
\nonumber\\
&&-\ep^2 s\(4t^2+2u^2+5tu\) +s^2t\ep^3\Big] {\rm Box}^6(s,t) \\
f_2(s,t,u)&=& 8(1-2\ep)\frac{t}{us^2}
\Big[(t-u)\(t^2+2u^2+tu\)+\ep\(-2t^3+ut^2+4tu^2+5u^3\)
\nonumber\\
&&-s^3\ep^2\Big] {\rm Box}^6(s,u) \\
f_3(s,t,u)&=& -4(1-\ep)\frac{t}{su} \bigg\{ 
\frac{V}{N}\lq 2s-t-\ep(2s-3u)-3s\ep^2\rq
 \nonumber \\
&& + 4N(u-u\ep+s\ep^2) \bigg\}\left[{\rm Bub}(u) - {\rm Bub}(s) \right]
\nonumber \\
&& + \frac{ 4t }{s^2u(1-\ep)(3-2\ep)} \left\{ -N\left[
18u^2+15t^2-3t(s-t) \right.\right.
\nonumber \\
&&\left.\left. -\ep\(78u^2-36t(s-t)+st\) +\ep^2\(80u^2+10s(s-t)-69st\)
\right]\right. 
\nonumber \\
&&\left. - \frac{1}{N}\left[-24u^2+3tu-21t^2  +\ep\(85u^2-43t(s-t)+3st\)
\right. \right. \nonumber \\ 
&& \left. \left.-\ep^2\(112u^2+6s(s-t)-109st\)
\right] 
 + \beta_0 \left[ 20s^2-40tu\right. \right. 
\nonumber \\
&& \left.\left. - 2\ep\(38s^2-3us-62tu\)+
4\ep^2\(27s^2-26tu\) \right] \right\} \ep{\rm Bub}(s) 
 +{\cal O}(\ep^3). 
\nonumber \\
\end{eqnarray}
These expressions are valid in all kinematic regions.  However, to evaluate
the pole structure in a particular region, the one-loop bubble graph ${\rm
Bub}$ 
and the one-loop box integral in $D=6-2\ep$ dimensions, ${\rm Box}^6$,  
must be expanded as a series
in $\epsilon$.  This analytic expansion is given in 
Appendix~\ref{app:master_int}.  

The function $\Htwo$, that appears in Eq.~(\ref{eq:poles}), 
exhibits only a single pole in $\ep$ and is given by 
\beq
\label{eq:htwo}
\Htwo \equiv \bra{\cm^{(0)}}{\bom H}^{(2)}(\ep)\ket{\cm^{(0)}} 
=\frac{e^{\ep \gamma}}{4\,\ep\,\Gamma(1-\ep)} H^{(2)} 
\braket{\cm^{(0)}}{\cm^{(0)}} \nonumber \\  
\eeq
where the constant $H^{(2)}$ is
\beqn
\label{eq:Htwo}
H^{(2)} &=&  
\left (\zeta_3+{\frac {5}{6}}+ {\frac {11}{72}}\,\pi^2
\right )\CA^2
+\left({13}\,\zeta_3+{\frac {245}{108}}-{\frac {23}{24}}\,{\pi 
}^{2}\right )\CA\,\CF+{\frac {10}{27}}\,\NF^2
\nonumber\\
&& +\left (-{\frac {{\pi }^{2}}{36}}-{
\frac {58}{27}}\right )\CA\,\NF+\left (\frac{{\pi }^{2}}{12}+{\frac
{29}{54}}\right )\NF\, \CF+
\left (-\frac{3}{4}+{\pi^2}-12\,\zeta_3\right )\CF^2, 
\nonumber \\   
\eeqn
and $\zeta_n$ is the Riemann Zeta function with $\zeta_2 = \pi^2/6$ and
$\zeta_3 = 1.202056\ldots$  We note that $H^{(2)}$ is renormalisation-scheme
dependent and Eq.~(\ref{eq:Htwo}) is valid in the \MSbar\ scheme.  We also
note that Eq.~(\ref{eq:Htwo}) differs from the corresponding expressions
found in the singularity structure of two-loop 
quark-quark scattering in all but the
$\CF^2$ coefficient.  This is due to the presence of infrared emissions from
gluons which modify the terms involving either $\CA$ or $\NF$.

It can be easily noted that the leading infrared singularity in 
Eq.~(\ref{eq:poles}) is $\O{1/\ep^4}$.  It is a very stringent check on the
reliability of our calculation that  the pole structure obtained by computing
the Feynman diagrams directly and introducing series expansions in $\epsilon$
for the scalar master integrals  agrees with Eq.~(\ref{eq:poles}) through to
$\O{1/\ep}$.   We therefore construct the finite remainder by subtracting
Eq.~(\ref{eq:poles}) from the full result.

\subsection{Finite contributions}
\label{subsec:finitetwo}

The finite two-loop contribution to $\C^8(s,t,u)$ is defined as 
\beq
\Finite(s,t,u) = 2 \Re \left[\braket{\cm^{(0)}}{\cm^{(2)\, {\rm fin}}}\right].
\eeq
In hadronic collisions, all parton scattering processes
(Eqs.~(\ref{eq:qqgg})--(\ref{eq:qgqg})) contribute simultaneously. We
therefore need to evaluate 
$\Finite(s,t,u)$ for the $q \bar q \to gg$  and $gg \to q \bar q$ process 
(which we denote as the $s$-channel since, although the tree-level process
contains graphs in all three channels, the squared tree matrix elements 
are proportional
to $1/s^2$) and
$\Finite(u,t,s)$ for the QCD Compton processes $q g \to q g$ and $g \bar q \to
g \bar q$ (which we label as the $u$-channel).  

Of course, the analytic expressions for the various processes
are related by crossing symmetry.  However, the master crossed boxes
have cuts in all three channels yielding complex parts in all physical
regions.  The analytic continuation is therefore rather involved and prone to
error.  We therefore choose to give expressions describing $\C^8(s,t,u)$ and
$\C^8(u,t,s)$ which are directly valid in the physical region $s > 0$ and $u,
t < 0$, and are given in terms of logarithms and polylogarithms that have no
imaginary parts.

As usual, the polylogarithms ${\rm Li}_n(w)$ are defined by
\begin{eqnarray}
 {\rm Li}_n(w) &=& \int_0^w \frac{dt}{t} {\rm Li}_{n-1}(t) \qquad {\rm ~for~}
 n=2,3,4\nonumber \\
 {\rm Li}_2(w) &=& -\int_0^w \frac{dt}{t} \log(1-t).
\label{eq:lidef}
\end{eqnarray} 
Using the standard polylogarithm identities~\cite{kolbig},
we retain the polylogarithms with arguments $x$, $1-x$ and
$(x-1)/x$, where
\begin{equation}
\label{eq:xydef}
x = -\frac{t}{s}, \qquad y = -\frac{u}{s} = 1-x, \qquad z=-\frac{u}{t} = \frac{x-1}{x}.
\end{equation}
For convenience, we also introduce the following logarithms
\begin{equation}
\label{eq:xydef1}
\lnx = \log\left(\frac{-t}{s}\right),
\qquad \lny = \log\left(\frac{-u}{s}\right),
\qquad \Ls = \log\left(\frac{s}{\mu^2}\right),
\qquad \Lu = \log\left(\frac{-u}{\mu^2}\right),
\end{equation}
where $\mu$ is the renormalisation scale.

For each channel, we choose to present our results by grouping terms 
according to the
power of the number of colours $N$ and the number 
of light quarks $\NF$, so that in the generic $c$-channel we write \\
\beqn
\label{eq:zi}
&&\Finite_c(s,t,u) = \(N^2-1\) \nonumber \\
&&\phantom{~}\times
\Bigg(
N^3 A_c + N B_c + \frac{1}{N} C_c  + \frac{1}{N^3} D_c  
+ \NF N^2 E_c + \NF F_c +  \frac{\NF}{N^2} G_c
+ \NF^2 N H_c + \frac{\NF^2}{N} I_c  
\Bigg).\nonumber \\
\eeqn

\subsubsection{$\boldsymbol{\Finite(s,t,u)}$: the $\boldsymbol{s}$-channel
process} 
\label{subsec:uttex}
We first give expressions valid for the annihilation processes, $q \qb
\to gg$ and $gg \to \bar q q$.  We find that
\begin{eqnarray}
{A_s }&=&{ \Biggl\{}\Biggl ({}-{36077\over 864}-{109\over 144}\,{\pi^4}+{241\over 12}\,{\zeta_3}-{5\over 2}\,{\pi^2}+{2327\over 216}\,{\Ls}+{20}\,{\Lidx}+{11\over 72}\,{\pi^2}\,{\Ls}-{9}\,{\zeta_3}\,{\Ls}\nonumber \\ &&
+{121\over 9}\,{\Ls^2}-{21}\,{\Lidy}-{20}\,{\Lidz}-{35\over 6}\,{\Licy}+{1\over 3}\,{\Lx}\,{\Ly^3}+{35\over 12}\,{\Ly^2}\,{\Lx}+{10\over 3}\,{\Ly}\,{\Lx}\,{\pi^2}\nonumber \\ &&
+{4}\,{\pi^2}\,{\Libx}+{16}\,{\Ly}\,{\zeta_3}+{1\over 2}\,{\Ly^2}\,{\Liby}+{35\over 6}\,{\Ly}\,{\Liby}+{121\over 9}\,{\Ly}\,{\Ls}-{5}\,{\Lx^2}\,{\Ly^2}\nonumber \\ &&
+{505\over 72}\,{\Ly^2}-{5}\,{\Lx^2}-{77\over 36}\,{\Ly^3}+{1\over 8}\,{\Ly^4}-{5\over 6}\,{\Lx^4}+{10\over 3}\,{\Ly}\,{\Lx^3}-{2\over 3}\,{\pi^2}\,{\Ly^2}-{5\over 3}\,{\pi^2}\,{\Lx^2}\nonumber \\ &&
-{143\over 36}\,{\Ly}\,{\pi^2}-{11\over 3}\,{\Ly^2}\,{\Ls}-{20}\,{\Ly}\,{\Licx}+{2273\over 216}\,{\Ly}\Biggr ){}\,{\sstt } \nonumber \\ &&
+{\utss}\,{}\Biggl ({11\over 90}\,{\pi^4}+{9\over 2}\,{\zeta_3}+{169\over 36}\,{\pi^2}+{457\over 144}-{11\over 6}\,{\Ls}+{33\over 4}\,{\Lx}\,{\Libx}-{8}\,{\Ly}\,{\zeta_3}+{33\over 8}\,{\Lx^2}\,{\Ly}\nonumber \\ &&
-{33\over 4}\,{\Licx}-{2417\over 144}\,{\Lx}+{2\over 3}\,{\Ly}\,{\pi^2}+{8}\,{\Ly}\,{\Licx}-{17\over 24}\,{\Lx^3}+{141\over 16}\,{\Ly^2}+{55\over 6}\,{\Ly}\,{\Ls}+{\Lx^2}\,{\Ly^2}\Biggr ){}\nonumber \\ &&
+{}\Biggl ({29\over 72}\,{\pi^4}-{205\over 6}\,{\zeta_3}-{2}\,{\pi^2}-{1535\over 108}\,{\Ls}+{30377\over 432}-{22}\,{\Lidx}-{11\over 36}\,{\pi^2}\,{\Ls}+{18}\,{\zeta_3}\,{\Ls}\nonumber \\ &&
-{242\over 9}\,{\Ls^2}+{18}\,{\Lidy}+{16}\,{\Lidz}+{5\over 12}\,{\Licy}+{57\over 4}\,{\Licx}-{2\over 3}\,{\Lx}\,{\Ly^3}-{5\over 24}\,{\Ly^2}\,{\Lx}\nonumber \\ &&
-{8\over 3}\,{\Ly}\,{\Lx}\,{\pi^2}+{977\over 144}\,{\Lx}-{4}\,{\pi^2}\,{\Libx}-{\Ly^2}\,{\Liby}-{5\over 12}\,{\Ly}\,{\Liby}-{55\over 6}\,{\Lx}\,{\Ls}\nonumber \\ &&
+{22\over 3}\,{\Ls}\,{\Lx^2}-{8}\,{\Licy}\,{\Lx}-{715\over 18}\,{\Ly}\,{\Ls}+{2}\,{\Lx^2}\,{\Ly^2}-{57\over 4}\,{\Lx}\,{\Libx}-{57\over 8}\,{\Lx^2}\,{\Ly}\nonumber \\ &&
+{4}\,{\Lx}\,{\Licx}+{4}\,{\Lx}\,{\zeta_3}-{\Lx^2}\,{\Libx}+{40\over 9}\,{\Lx}\,{\pi^2}-{3847\over 144}\,{\Ly^2}-{793\over 144}\,{\Lx^2}+{401\over 72}\,{\Ly^3}\nonumber \\ &&
-{1\over 4}\,{\Ly^4}+{1\over 12}\,{\Lx^4}+{431\over 72}\,{\Lx^3}-{8\over 3}\,{\Ly}\,{\Lx^3}+{4\over 3}\,{\pi^2}\,{\Ly^2}+{4}\,{\pi^2}\,{\Lx^2}+{47\over 18}\,{\Ly}\,{\pi^2}+{22\over 3}\,{\Ly^2}\,{\Ls}\nonumber \\ &&
+{8}\,{\Ly}\,{\Licx}-{3059\over 432}\,{\Ly}\Biggr ){}\,{\tttu } \Biggr \} + \Biggl \{ u \leftrightarrow t \Biggr \}
\end{eqnarray}

\begin{eqnarray}
{B_s }&=&{\Biggl \{ }\Biggl ({317\over 180}\,{\pi^4}-{733\over 18}\,{\zeta_3}-{1055\over 72}\,{\pi^2}-{7543\over 216}\,{\Ls}+{2}\,{\Ly}\,{\Licy}+{8}\,{\Lidx}-{20}\,{\Ly}\,{\Lx}\,{\Liby}\nonumber \\ &&
-{44\over 3}\,{\Ly}\,{\Lx}\,{\Ls}+{65\over 9}\,{\pi^2}\,{\Ls}+{10}\,{\zeta_3}\,{\Ls}-{121\over 9}\,{\Ls^2}-{79}\,{\Lidy}-{74}\,{\Lidz}-{71\over 6}\,{\Licy}\nonumber \\ &&
+{25\over 3}\,{\Licx}+{164771\over 2592}+{5\over 3}\,{\Lx}\,{\Ly^3}+{1\over 4}\,{\Ly^2}\,{\Lx}+{19\over 3}\,{\Ly}\,{\Lx}\,{\pi^2}-{22\over 27}\,{\Lx}-{19\over 3}\,{\pi^2}\,{\Libx}\nonumber \\ &&
+{74}\,{\Ly}\,{\zeta_3}+{1\over 2}\,{\Ly^2}\,{\Liby}+{71\over 6}\,{\Ly}\,{\Liby}+{16\over 9}\,{\Lx}\,{\Ly}-{220\over 9}\,{\Lx}\,{\Ls}+{24}\,{\Licy}\,{\Lx}\nonumber \\ &&
-{22\over 9}\,{\Ly}\,{\Ls}-{49\over 2}\,{\Lx^2}\,{\Ly^2}-{25\over 3}\,{\Lx}\,{\Libx}+{55\over 6}\,{\Lx^2}\,{\Ly}+{44}\,{\Lx}\,{\Licx}-{36}\,{\Lx}\,{\zeta_3}\nonumber \\ &&
-{11}\,{\Lx^2}\,{\Libx}-{551\over 36}\,{\Lx}\,{\pi^2}-{125\over 72}\,{\Ly^2}+{583\over 36}\,{\Lx^2}+{127\over 36}\,{\Ly^3}-{3\over 8}\,{\Ly^4}-{25\over 12}\,{\Lx^4}-{17\over 2}\,{\Lx^3}\nonumber \\ &&
+{9}\,{\Ly}\,{\Lx^3}-{8\over 3}\,{\pi^2}\,{\Ly^2}-{16\over 3}\,{\pi^2}\,{\Lx^2}-{83\over 36}\,{\Ly}\,{\pi^2}+{22\over 3}\,{\Ly^2}\,{\Ls}-{70}\,{\Ly}\,{\Licx}-{5765\over 216}\,{\Ly}\Biggr ){}\,{\sstt } \nonumber \\ &&
+{\utss}\,{}\Biggl ({}-{229\over 240}\,{\pi^4}+{118\over 3}\,{\zeta_3}-{799\over 72}\,{\pi^2}-{11\over 6}\,{\Ls}+{139\over 36}-{18}\,{\Ly}\,{\Licy}+{65\over 2}\,{\Lidy}\nonumber \\ &&
-{299\over 6}\,{\Licx}+{2\over 3}\,{\Lx}\,{\Ly^3}-{5\over 6}\,{\Ly}\,{\Lx}\,{\pi^2}+{14\over 9}\,{\Lx}+{19}\,{\Ly}\,{\zeta_3}+{7\over 4}\,{\Ly^2}\,{\Liby}+{73\over 4}\,{\Lx}\,{\Ly}\nonumber \\ &&
-{55\over 6}\,{\Ly}\,{\Ls}-{13\over 8}\,{\Lx^2}\,{\Ly^2}+{299\over 6}\,{\Lx}\,{\Libx}+{62\over 3}\,{\Lx^2}\,{\Ly}-{311\over 36}\,{\Ly^2}-{17\over 48}\,{\Ly^4}+{73\over 18}\,{\Lx^3}\nonumber \\ &&
-{47\over 12}\,{\pi^2}\,{\Ly^2}-{55\over 6}\,{\Ly}\,{\pi^2}-{11\over 6}\,{\Ly^2}\,{\Ls}-{9}\,{\Ly}\,{\Licx}\Biggr ){}\nonumber \\ &&
+{}\Biggl ({}-{17\over 60}\,{\pi^4}+{443\over 18}\,{\zeta_3}+{16\over 9}\,{\pi^2}+{1502\over 27}\,{\Ls}+{20}\,{\Ly}\,{\Licy}+{7\over 2}\,{\Lidx}+{8}\,{\Ly}\,{\Lx}\,{\Liby}\nonumber \\ &&
-{1\over 12}\,{\pi^2}\,{\Ls}-{2}\,{\zeta_3}\,{\Ls}-{79\over 2}\,{\Lidy}-{22}\,{\Lidz}+{71\over 6}\,{\Licy}-{71\over 6}\,{\Licx}-{16\over 3}\,{\Lx}\,{\Ly^3}\nonumber \\ &&
+{32\over 3}\,{\Ly^2}\,{\Lx}-{8\over 3}\,{\Ly}\,{\Lx}\,{\pi^2}+{359\over 36}\,{\Lx}+{7}\,{\pi^2}\,{\Libx}-{\Ly}\,{\zeta_3}-{45\over 4}\,{\Ly^2}\,{\Liby}-{71\over 6}\,{\Ly}\,{\Liby}\nonumber \\ &&
+{391\over 18}\,{\Lx}\,{\Ly}+{11\over 6}\,{\Lx}\,{\Ls}+{11\over 6}\,{\Ls}\,{\Lx^2}+{7}\,{\Licy}\,{\Lx}-{11\over 6}\,{\Ly}\,{\Ls}+{13\over 4}\,{\Lx^2}\,{\Ly^2}+{71\over 6}\,{\Lx}\,{\Libx}\nonumber \\ &&
-{24}\,{\Lx^2}\,{\Ly}+{16}\,{\Lx}\,{\Licx}+{\Lx}\,{\zeta_3}-{27\over 4}\,{\Lx^2}\,{\Libx}+{103\over 18}\,{\Lx}\,{\pi^2}-{203\over 36}\,{\Ly^2}-{277\over 12}\,{\Lx^2}\nonumber \\ &&
-{101\over 18}\,{\Ly^3}+{31\over 48}\,{\Ly^4}-{33\over 16}\,{\Lx^4}+{341\over 18}\,{\Lx^3}+{3}\,{\Ly}\,{\Lx^3}+{11\over 4}\,{\pi^2}\,{\Ly^2}+{11\over 4}\,{\pi^2}\,{\Lx^2}-{103\over 18}\,{\Ly}\,{\pi^2}\nonumber \\ &&
-{11\over 6}\,{\Ly^2}\,{\Ls}-{7}\,{\Ly}\,{\Licx}-{19139\over 324}+{601\over 36}\,{\Ly}\Biggr ){}\,{\tttu } \nonumber \\ &&
-{\Lx^2}\,{\tfiveou } \Biggr \} + \Biggl \{ u \leftrightarrow t \Biggr \}
\end{eqnarray}
\begin{eqnarray}
{C_s }&=&{\Biggl \{ }\Biggl ({}-{41393\over 2592}+{301\over 720}\,{\pi^4}+{173\over 36}\,{\zeta_3}-{133\over 36}\,{\pi^2}+{5297\over 216}\,{\Ls}+{4}\,{\Ly}\,{\Licy}-{20}\,{\Lidx}\nonumber \\ &&
-{4}\,{\Ly}\,{\Lx}\,{\Liby}-{13\over 24}\,{\pi^2}\,{\Ls}+{5}\,{\zeta_3}\,{\Ls}+{\Lidy}+{8}\,{\Lidz}-{15\over 2}\,{\Licy}-{3}\,{\Licx}\nonumber \\ &&
-{2\over 3}\,{\Lx}\,{\Ly^3}-{11\over 12}\,{\Ly^2}\,{\Lx}-{14\over 3}\,{\Ly}\,{\Lx}\,{\pi^2}+{59\over 3}\,{\Lx}-{8\over 3}\,{\pi^2}\,{\Libx}-{12}\,{\Ly}\,{\zeta_3}-{1\over 2}\,{\Ly^2}\,{\Liby}\nonumber \\ &&
+{15\over 2}\,{\Ly}\,{\Liby}+{2}\,{\Lx}\,{\Ly}-{22\over 3}\,{\Ls}\,{\Lx^2}+{10}\,{\Licy}\,{\Lx}-{11}\,{\Ly}\,{\Ls}+{7\over 2}\,{\Lx^2}\,{\Ly^2}+{3}\,{\Lx}\,{\Libx}\nonumber \\ &&
-{10\over 3}\,{\Lx^2}\,{\Ly}+{10}\,{\Lx}\,{\Licx}-{6}\,{\Lx}\,{\zeta_3}-{4}\,{\Lx^2}\,{\Libx}+{49\over 18}\,{\Lx}\,{\pi^2}-{425\over 72}\,{\Ly^2}+{179\over 18}\,{\Lx^2}\nonumber \\ &&
-{23\over 36}\,{\Ly^3}+{3\over 8}\,{\Ly^4}+{1\over 3}\,{\Lx^4}-{65\over 18}\,{\Lx^3}-{8\over 3}\,{\Ly}\,{\Lx^3}+{5\over 6}\,{\pi^2}\,{\Ly^2}+{3\over 2}\,{\pi^2}\,{\Lx^2}+{5\over 18}\,{\Ly}\,{\pi^2}\nonumber \\ &&
-{11\over 3}\,{\Ly^2}\,{\Ls}+{14}\,{\Ly}\,{\Licx}+{271\over 24}\,{\Ly}\Biggr ){}\,{\sstt } \nonumber \\ &&
+{\utss}\,{}\Biggl ({271\over 240}\,{\pi^4}+{3\over 2}\,{\zeta_3}-{127\over 72}\,{\pi^2}+{13\over 4}\,{\Lx}\,{\Libx}+{26}\,{\Ly}\,{\Licy}-{21\over 4}\,{\Ly^2}\,{\Liby}-{25}\,{\Ly}\,{\zeta_3}\nonumber \\ &&
-{119\over 24}\,{\Lx^2}\,{\Ly}-{13\over 4}\,{\Licx}+{3133\over 144}\,{\Lx}-{4\over 3}\,{\Lx}\,{\Ly^3}+{7\over 4}\,{\pi^2}\,{\Ly^2}+{55\over 18}\,{\Ly}\,{\pi^2}-{55\over 12}\,{\Lx}\,{\Ly}\nonumber \\ &&
-{83\over 2}\,{\Lidy}-{7\over 6}\,{\Ly}\,{\Lx}\,{\pi^2}-{22\over 3}\,{\Ly^2}\,{\Ls}+{15}\,{\Ly}\,{\Licx}+{23\over 16}-{209\over 72}\,{\Lx^3}+{919\over 144}\,{\Ly^2}\nonumber \\ &&
+{9\over 16}\,{\Ly^4}-{22\over 3}\,{\Ly}\,{\Ls}+{23\over 8}\,{\Lx^2}\,{\Ly^2}\Biggr ){}\nonumber \\ &&
+{}\Biggl ({13\over 360}\,{\pi^4}+{15}\,{\zeta_3}+{1\over 3}\,{\pi^2}-{3\over 4}\,{\Ls}+{11\over 2}\,{\Lidx}+{\pi^2}\,{\Ls}-{12}\,{\zeta_3}\,{\Ls}-{11\over 2}\,{\Lidy}-{6}\,{\Lidz}\nonumber \\ &&
+{11\over 4}\,{\Licy}-{11\over 4}\,{\Licx}-{255\over 16}-{25\over 8}\,{\Ly^2}\,{\Lx}+{2\over 3}\,{\Ly}\,{\Lx}\,{\pi^2}-{21\over 16}\,{\Lx}+{1\over 3}\,{\pi^2}\,{\Libx}\nonumber \\ &&
+{3}\,{\Ly}\,{\zeta_3}-{1\over 4}\,{\Ly^2}\,{\Liby}-{11\over 4}\,{\Ly}\,{\Liby}-{11\over 2}\,{\Lx}\,{\Ly}+{3}\,{\Licy}\,{\Lx}-{3\over 4}\,{\Lx^2}\,{\Ly^2}\nonumber \\ &&
+{11\over 4}\,{\Lx}\,{\Libx}+{25\over 8}\,{\Lx^2}\,{\Ly}-{3}\,{\Lx}\,{\zeta_3}+{1\over 4}\,{\Lx^2}\,{\Libx}-{7\over 6}\,{\Lx}\,{\pi^2}+{83\over 16}\,{\Ly^2}-{11\over 16}\,{\Lx^2}\nonumber \\ &&
+{13\over 24}\,{\Ly^3}-{5\over 48}\,{\Ly^4}-{7\over 48}\,{\Lx^4}-{13\over 24}\,{\Lx^3}+{\Ly}\,{\Lx^3}+{1\over 12}\,{\pi^2}\,{\Ly^2}-{7\over 12}\,{\pi^2}\,{\Lx^2}+{7\over 6}\,{\Ly}\,{\pi^2}\nonumber \\ &&
-{3}\,{\Ly}\,{\Licx}+{21\over 16}\,{\Ly}\Biggr ){}\,{\tttu } 
+{\Lx^2}\,{\tfiveou } \Biggr \} + \Biggl \{ u \leftrightarrow t \Biggr \}
\end{eqnarray}

\begin{eqnarray}
{D_s }&=&{\Biggl \{ }\Biggl ({}-{1\over 90}\,{\pi^4}+{5\over 2}\,{\zeta_3}+{5\over 8}\,{\pi^2}-{3\over 8}\,{\Ls}+{2}\,{\Ly}\,{\Licy}+{1\over 2}\,{\pi^2}\,{\Ls}-{6}\,{\zeta_3}\,{\Ls}+{11}\,{\Lidy}\nonumber \\ &&
+{14}\,{\Lidz}-{3\over 2}\,{\Licy}+{8}\,{\Licx}-{187\over 32}+{3\over 4}\,{\Ly^2}\,{\Lx}-{7\over 3}\,{\Ly}\,{\Lx}\,{\pi^2}+{\Lx}+{\pi^2}\,{\Libx}\nonumber \\ &&
-{18}\,{\Ly}\,{\zeta_3}-{1\over 2}\,{\Ly^2}\,{\Liby}+{3\over 2}\,{\Ly}\,{\Liby}+{2}\,{\Licy}\,{\Lx}+{4}\,{\Lx^2}\,{\Ly^2}-{8}\,{\Lx}\,{\Libx}\nonumber \\ &&
-{11\over 2}\,{\Lx^2}\,{\Ly}-{6}\,{\Lx}\,{\Licx}+{4}\,{\Lx}\,{\zeta_3}-{\Lx^2}\,{\Libx}+{10\over 3}\,{\Lx}\,{\pi^2}+{5\over 8}\,{\Ly^2}-{9\over 4}\,{\Lx^2}-{3\over 4}\,{\Ly^3}\nonumber \\ &&
-{1\over 8}\,{\Ly^4}-{1\over 12}\,{\Lx^4}-{7\over 3}\,{\Ly}\,{\Lx^3}+{1\over 2}\,{\pi^2}\,{\Ly^2}+{1\over 2}\,{\pi^2}\,{\Lx^2}+{2}\,{\Ly}\,{\pi^2}+{16}\,{\Ly}\,{\Licx}+{39\over 8}\,{\Ly}\Biggr ){}\,{\sstt } \nonumber \\ &&
+{\utss}\,{}\Biggl ({1\over 4}-{4\over 45}\,{\pi^4}-{2}\,{\zeta_3}+{43\over 12}\,{\pi^2}-{19}\,{\Lx}\,{\Libx}-{4}\,{\Ly}\,{\Licy}-{\Ly^2}\,{\Liby}-{12}\,{\Ly}\,{\zeta_3}\nonumber \\ &&
-{15\over 2}\,{\Lx^2}\,{\Ly}+{19}\,{\Licx}+{29\over 4}\,{\Lx}-{1\over 3}\,{\pi^2}\,{\Ly^2}+{4\over 3}\,{\Ly}\,{\pi^2}-{7\over 2}\,{\Lx}\,{\Ly}+{10}\,{\Lidy}\nonumber \\ &&
-{4\over 3}\,{\Ly}\,{\Lx}\,{\pi^2}+{16}\,{\Ly}\,{\Licx}-{11\over 6}\,{\Lx^3}+{4}\,{\Ly^2}-{7\over 12}\,{\Ly^4}+{2}\,{\Lx^2}\,{\Ly^2}\Biggr ){}\nonumber \\ &&
+{}\Biggl ({}-{3}\,{\Lx}+{7\over 2}\,{\Lx^2}+{3}\,{\Ly}+{3\over 2}\,{\pi^2}+{1\over 2}\,{\Ly^2}-{3}\,{\Lx}\,{\Ly}\Biggr ){}\,{\tttu } \nonumber \\ &&-{\Lx^2}\,{\tfiveou } \Biggr \} + \Biggl \{ u \leftrightarrow t \Biggr \}
\end{eqnarray}
\begin{eqnarray}
{E_s }&=&{\Biggl \{ }\Biggl ({14\over 45}\,{\pi^4}-{17\over 6}\,{\zeta_3}+{77\over 54}\,{\pi^2}+{185\over 54}\,{\Ls}-{8}\,{\Lidx}-{1\over 36}\,{\pi^2}\,{\Ls}-{44\over 9}\,{\Ls^2}+{8}\,{\Lidy}\nonumber \\ &&
+{8}\,{\Lidz}+{4\over 3}\,{\Licy}-{2\over 3}\,{\Ly^2}\,{\Lx}-{4\over 3}\,{\Ly}\,{\Lx}\,{\pi^2}-{4\over 3}\,{\pi^2}\,{\Libx}-{8}\,{\Ly}\,{\zeta_3}-{4\over 3}\,{\Ly}\,{\Liby}\nonumber \\ &&
-{44\over 9}\,{\Ly}\,{\Ls}+{2}\,{\Lx^2}\,{\Ly^2}-{43\over 18}\,{\Ly^2}+{2}\,{\Lx^2}+{7\over 18}\,{\Ly^3}+{1\over 3}\,{\Lx^4}-{4\over 3}\,{\Ly}\,{\Lx^3}+{2\over 3}\,{\pi^2}\,{\Lx^2}+{13\over 18}\,{\Ly}\,{\pi^2}\nonumber \\ &&
+{2\over 3}\,{\Ly^2}\,{\Ls}+{8}\,{\Ly}\,{\Licx}+{10\over 9}\,{\Ly}+{1307\over 216}\Biggr ){}\,{\sstt } \nonumber \\ &&
+{\utss}\,{}\Biggl ({}-{37\over 36}-{11\over 360}\,{\pi^4}-{4}\,{\zeta_3}-{49\over 36}\,{\pi^2}+{1\over 3}\,{\Ls}-{4}\,{\Lx}\,{\Libx}+{2}\,{\Ly}\,{\zeta_3}-{2}\,{\Lx^2}\,{\Ly}+{4}\,{\Licx}\nonumber \\ &&
+{91\over 18}\,{\Lx}+{2\over 3}\,{\Ly}\,{\pi^2}-{2}\,{\Ly}\,{\Licx}-{2}\,{\Ly^2}-{5\over 3}\,{\Ly}\,{\Ls}-{1\over 4}\,{\Lx^2}\,{\Ly^2}\Biggr ){}\nonumber \\ &&
+{}\Biggl ({}-{17\over 180}\,{\pi^4}+{17\over 3}\,{\zeta_3}-{32\over 27}\,{\pi^2}-{221\over 27}\,{\Ls}+{4}\,{\Lidx}+{1\over 18}\,{\pi^2}\,{\Ls}+{88\over 9}\,{\Ls^2}-{4}\,{\Lidy}\nonumber \\ &&
-{4}\,{\Lidz}+{4\over 3}\,{\Licy}-{4}\,{\Licx}-{2\over 3}\,{\Ly^2}\,{\Lx}+{2\over 3}\,{\Ly}\,{\Lx}\,{\pi^2}-{19\over 18}\,{\Lx}+{2\over 3}\,{\pi^2}\,{\Libx}\nonumber \\ &&
+{2}\,{\Ly}\,{\zeta_3}-{4\over 3}\,{\Ly}\,{\Liby}-{863\over 108}+{5\over 3}\,{\Lx}\,{\Ls}-{4\over 3}\,{\Ls}\,{\Lx^2}+{2}\,{\Licy}\,{\Lx}+{109\over 9}\,{\Ly}\,{\Ls}\nonumber \\ &&
-{1\over 2}\,{\Lx^2}\,{\Ly^2}+{4}\,{\Lx}\,{\Libx}+{2}\,{\Lx^2}\,{\Ly}-{2}\,{\Lx}\,{\zeta_3}-{16\over 9}\,{\Lx}\,{\pi^2}+{70\over 9}\,{\Ly^2}+{1\over 9}\,{\Lx^2}-{7\over 9}\,{\Ly^3}\nonumber \\ &&
-{1\over 6}\,{\Lx^4}-{7\over 9}\,{\Lx^3}+{2\over 3}\,{\Ly}\,{\Lx^3}-{1\over
3}\,{\pi^2}\,{\Lx^2}-{7\over 9}\,{\Ly}\,{\pi^2}-{4\over
3}\,{\Ly^2}\,{\Ls}-{2}\,{\Ly}\,{\Licx}-{31\over 6}\,{\Ly}\Biggr ){}\,{\tttu }
\Biggr \}\nonumber \\
&& + \Biggl \{ u \leftrightarrow t \Biggr \}
\end{eqnarray}

\begin{eqnarray}
{F_s}&=&{\Biggl \{ }\Biggl ({}-{28\over 45}\,{\pi^4}-{10\over 9}\,{\zeta_3}+{347\over 108}\,{\pi^2}+{7\over 9}\,{\Ls}+{16}\,{\Lidx}+{8\over 3}\,{\Ly}\,{\Lx}\,{\Ls}-{11\over 9}\,{\pi^2}\,{\Ls}+{44\over 9}\,{\Ls^2}\nonumber \\ &&
-{16}\,{\Lidy}-{16}\,{\Lidz}+{4\over 3}\,{\Licy}+{4}\,{\Licx}+{1\over 2}\,{\Ly^2}\,{\Lx}+{8\over 3}\,{\Ly}\,{\Lx}\,{\pi^2}-{161\over 18}\,{\Lx}\nonumber \\ &&
+{8\over 3}\,{\pi^2}\,{\Libx}+{16}\,{\Ly}\,{\zeta_3}-{4\over 3}\,{\Ly}\,{\Liby}-{1\over 2}\,{\Lx}\,{\Ly}-{3661\over 324}+{62\over 9}\,{\Lx}\,{\Ls}+{26\over 9}\,{\Ly}\,{\Ls}\nonumber \\ &&
-{4}\,{\Lx^2}\,{\Ly^2}-{4}\,{\Lx}\,{\Libx}-{2\over 3}\,{\Lx^2}\,{\Ly}-{37\over 18}\,{\Lx}\,{\pi^2}+{35\over 18}\,{\Ly^2}-{79\over 18}\,{\Lx^2}-{7\over 9}\,{\Ly^3}-{2\over 3}\,{\Lx^4}\nonumber \\ &&
+{2\over 3}\,{\Lx^3}+{8\over 3}\,{\Ly}\,{\Lx^3}-{4\over 3}\,{\pi^2}\,{\Lx^2}+{1\over 18}\,{\Ly}\,{\pi^2}-{4\over 3}\,{\Ly^2}\,{\Ls}-{16}\,{\Ly}\,{\Licx}+{23\over 9}\,{\Ly}\Biggr ){}\,{\sstt } \nonumber \\ &&
+{\utss}\,{}\Biggl ({}-{11\over 90}\,{\pi^4}+{52\over 3}\,{\zeta_3}+{35\over 12}\,{\pi^2}+{1\over 3}\,{\Ls}+{52\over 3}\,{\Lx}\,{\Libx}+{8}\,{\Ly}\,{\zeta_3}+{47\over 6}\,{\Lx^2}\,{\Ly}-{52\over 3}\,{\Licx}\nonumber \\ &&
-{50\over 9}\,{\Lx}+{1\over 18}\,{\Ly}\,{\pi^2}-{1\over 3}\,{\Lx}\,{\Ly}+{1\over 3}\,{\Ly^2}\,{\Ls}-{8}\,{\Ly}\,{\Licx}-{11\over 36}\,{\Lx^3}+{49\over 18}\,{\Ly^2}-{19\over 36}\nonumber \\ &&
+{5\over 3}\,{\Ly}\,{\Ls}-{\Lx^2}\,{\Ly^2}\Biggr ){}\nonumber \\ &&
+{}\Biggl ({}-{263\over 27}\,{\Ls}-{1\over 3}\,{\Lx}\,{\Ls}+{3\over 2}\,{\Lx}\,{\pi^2}-{1\over 3}\,{\Ls}\,{\Lx^2}+{5\over 2}\,{\Lx^2}\,{\Ly}+{1\over 3}\,{\Ly}\,{\Ls}-{1\over 6}\,{\pi^2}\,{\Ls}\nonumber \\ &&
+{1\over 3}\,{\Ly^2}\,{\Ls}+{7\over 9}\,{\Lx}+{4\over 3}\,{\Lx}\,{\Libx}+{211\over 18}\,{\Lx^2}-{4\over 3}\,{\Licx}+{4085\over 324}+{4\over 3}\,{\Licy}-{1\over 9}\,{\zeta_3}\nonumber \\ &&
-{31\over 9}\,{\Ly}-{4\over 3}\,{\Ly}\,{\Liby}-{3\over 2}\,{\Ly}\,{\pi^2}-{73\over 36}\,{\Lx^3}-{7\over 6}\,{\Ly^2}\,{\Lx}+{25\over 36}\,{\Ly^3}+{25\over 18}\,{\pi^2}-{41\over 18}\,{\Ly^2}\nonumber \\ &&
-{58\over 9}\,{\Lx}\,{\Ly}\Biggr ){}\,{\tttu } -{3}\,{\Lx^2}\,{\tfiveou } \Biggr \} + \Biggl \{ u \leftrightarrow t \Biggr \}
\end{eqnarray}

\begin{eqnarray}
{G_s }&=&{\Biggl \{ }\Biggl ({28\over 15}\,{\pi^4}+{143\over 18}\,{\zeta_3}+{19\over 36}\,{\pi^2}-{227\over 54}\,{\Ls}-{48}\,{\Lidx}-{1\over 12}\,{\pi^2}\,{\Ls}+{48}\,{\Lidy}+{48}\,{\Lidz}\nonumber \\ &&
-{8}\,{\Licx}+{3401\over 648}+{1\over 6}\,{\Ly^2}\,{\Lx}-{8}\,{\Ly}\,{\Lx}\,{\pi^2}+{101\over 6}\,{\Lx}-{8}\,{\pi^2}\,{\Libx}-{48}\,{\Ly}\,{\zeta_3}\nonumber \\ &&
+{1\over 2}\,{\Lx}\,{\Ly}+{4\over 3}\,{\Ls}\,{\Lx^2}+{2}\,{\Ly}\,{\Ls}+{12}\,{\Lx^2}\,{\Ly^2}+{8}\,{\Lx}\,{\Libx}+{1\over 3}\,{\Lx^2}\,{\Ly}+{64\over 9}\,{\Lx}\,{\pi^2}\nonumber \\ &&
+{4\over 9}\,{\Ly^2}+{89\over 9}\,{\Lx^2}+{7\over 18}\,{\Ly^3}+{2}\,{\Lx^4}+{7\over 9}\,{\Lx^3}-{8}\,{\Ly}\,{\Lx^3}+{4}\,{\pi^2}\,{\Lx^2}+{2\over 9}\,{\Ly}\,{\pi^2}+{2\over 3}\,{\Ly^2}\,{\Ls}\nonumber \\ &&
+{48}\,{\Ly}\,{\Licx}-{11\over 3}\,{\Ly}\Biggr ){}\,{\sstt } \nonumber \\ &&
+{\utss}\,{}\Biggl ({11\over 30}\,{\pi^4}-{52}\,{\zeta_3}-{107\over 18}\,{\pi^2}-{52}\,{\Lx}\,{\Libx}-{24}\,{\Ly}\,{\zeta_3}-{71\over 3}\,{\Lx^2}\,{\Ly}+{52}\,{\Licx}+{71\over 9}\,{\Lx}\nonumber \\ &&
+{46\over 9}\,{\Ly}\,{\pi^2}-{2\over 3}\,{\Lx}\,{\Ly}+{4\over 3}\,{\Ly^2}\,{\Ls}+{24}\,{\Ly}\,{\Licx}+{7\over 9}\,{\Lx^3}-{37\over 9}\,{\Ly^2}+{4\over 3}\,{\Ly}\,{\Ls}+{3}\,{\Lx^2}\,{\Ly^2}\Biggr ){}\nonumber \\ &&
+{}\Biggl ({}-{10\over 3}\,{\Lx}\,{\pi^2}-{8}\,{\Lx}-{4}\,{\Lx}\,{\Libx}-{18}\,{\Lx^2}+{4}\,{\Licx}-{4}\,{\Licy}+{8}\,{\Ly}+{4}\,{\Ly}\,{\Liby}\nonumber \\ &&
+{10\over 3}\,{\Ly}\,{\pi^2}+{9}\,{\Ly^2}\Biggr ){}\,{\tttu } +{9}\,{\Lx^2}\,{\tfiveou } \Biggr \} + \Biggl \{ u \leftrightarrow t \Biggr \}
\end{eqnarray}

\begin{eqnarray}
{H_s }&=&{\Biggl \{ }\Biggl ({}-{4\over 27}\,{\pi^2}+{4\over 9}\,{\Ls^2}-{20\over 27}\,{\Ls}-{10\over 27}\,{\Ly}+{4\over 9}\,{\Ly}\,{\Ls}+{1\over 6}\,{\Ly^2}\Biggr ){}\,{\sstt } \nonumber \\ &&
+{}\Biggl ({8\over 27}\,{\pi^2}+{40\over 27}\,{\Ls}-{8\over 9}\,{\Ls^2}+{20\over 27}\,{\Ly}-{8\over 9}\,{\Ly}\,{\Ls}-{1\over 3}\,{\Ly^2}\Biggr ){}\,{\tttu } \Biggr \} + \Biggl \{ u \leftrightarrow t \Biggr \}
\end{eqnarray}

\begin{eqnarray}
{I_s }&=&{\Biggl \{ }\Biggl ({}-{1\over 9}\,{\Lx}\,{\Ly}+{10\over 27}\,{\Ly}-{1\over 6}\,{\Ly^2}+{20\over 27}\,{\Ls}-{4\over 9}\,{\Lx}\,{\Ls}-{4\over 9}\,{\Ls^2}-{1\over 18}\,{\Lx^2}-{7\over 54}\,{\pi^2} \nonumber \\
&& -{4\over 9}\,{\Ly}\,{\Ls}+{10\over 27}\,{\Lx}\Biggr ){}\,{\sstt } -{2\over 9}\,{\Lx}\,{}\Biggl ({\Lx}-{\Ly}\Biggr ){}\,{\tttu } \Biggr \} + \Biggl \{ u \leftrightarrow t \Biggr \}\end{eqnarray}

\subsubsection{$\boldsymbol{\Finite(u,t,s)}$: the $\boldsymbol{u}$-channel
process} 
\label{subsec:sttex}
Similarly, for the Compton scattering processes,
$q g \to q g$ and $g \bar q \to
g \bar q$, we find that the
coefficients for the finite part in Eq.~(\ref{eq:zi}) are given by
\begin{eqnarray}
{A_u}&=&{ }\Biggl ({}-{2417\over 144}\,{\Lx}+{11\over 45}\,{\pi^4}-{19\over 144}\,{\pi^2}+{3\over 4}\,{\zeta_3}-{11\over 3}\,{\Lu}+{33\over 4}\,{\Licx}+{457\over 72}+{8}\,{\Licx}\,{\Ly}\nonumber \\ &&
-{55\over 3}\,{\Lu}\,{\Ly}-{10\over 3}\,{\pi^2}\,{\Ly}+{8}\,{\zeta_3}\,{\Ly}-{2\over 3}\,{\pi^2}\,{\Ly^2}-{2}\,{\Lx^2}\,{\Ly}-{33\over 4}\,{\Lx}\,{\Libx}-{8}\,{\Lx}\,{\zeta_3}\nonumber \\ &&
-{35\over 24}\,{\Lx}\,{\pi^2}+{4\over 3}\,{\Lx}\,{\Ly^3}+{8}\,{\Licy}\,{\Lx}+{55\over 6}\,{\Lu}\,{\Lx}+{141\over 16}\,{\Lx^2}-{17\over 24}\,{\Lx^3}-{141\over 8}\,{\Ly}\,{\Lx}\nonumber \\ &&
+{2}\,{\Ly^2}\,{\Lx}+{2}\,{\Lx^2}\,{\Ly^2}-{8\over 3}\,{\pi^2}\,{\Lx}\,{\Ly}-{2\over 3}\,{\Ly^4}-{4\over 3}\,{\Ly^3}+{141\over 8}\,{\Ly^2}+{2417\over 72}\,{\Ly}\Biggr ){}\,{\stuu } \nonumber \\ &&
+{}\Biggl ({121\over 9}\,{\Lu^2}-{36077\over 864}-{9}\,{\Lu}\,{\zeta_3}-{121\over 72}\,{\Lu}\,{\pi^2}+{2273\over 432}\,{\Lx}-{181\over 720}\,{\pi^4}-{215\over 144}\,{\pi^2}+{103\over 6}\,{\zeta_3}\nonumber \\ &&
+{2327\over 216}\,{\Lu}+{1\over 2}\,{\Lidz}+{1\over 2}\,{\Lidy}+{35\over 12}\,{\Licx}-{10}\,{\Licx}\,{\Ly}-{121\over 9}\,{\Lu}\,{\Ly}-{29\over 18}\,{\pi^2}\,{\Ly}\nonumber \\ &&
-{6}\,{\zeta_3}\,{\Ly}+{7\over 12}\,{\pi^2}\,{\Ly^2}+{7\over 4}\,{\Lx^2}\,{\Ly}-{35\over 12}\,{\Lx}\,{\Libx}+{8}\,{\Lx}\,{\zeta_3}-{1\over 4}\,{\Lx^2}\,{\Libx}-{55\over 36}\,{\Lx}\,{\pi^2}\nonumber \\ &&
-{11\over 6}\,{\Lu}\,{\Lx^2}-{13\over 6}\,{\Lx}\,{\Ly^3}+{1\over 12}\,{\pi^2}\,{\Lx^2}-{1\over 4}\,{\pi^2}\,{\Libx}-{11\over 3}\,{\Lu}\,{\Ly^2}-{10}\,{\Licy}\,{\Lx}\nonumber \\ &&
+{121\over 18}\,{\Lu}\,{\Lx}+{1\over 12}\,{\Lx^4}+{145\over 144}\,{\Lx^2}-{77\over 72}\,{\Lx^3}-{1\over 2}\,{\Lx^3}\,{\Ly}-{145\over 72}\,{\Ly}\,{\Lx}-{7\over 4}\,{\Ly^2}\,{\Lx}-{17\over 8}\,{\Lx^2}\,{\Ly^2}\nonumber \\ &&
+{37\over 12}\,{\pi^2}\,{\Lx}\,{\Ly}+{11\over 3}\,{\Lu}\,{\Lx}\,{\Ly}-{1\over 2}\,{\Liby}\,{\Lx}\,{\Ly}+{13\over 12}\,{\Ly^4}+{7\over 6}\,{\Ly^3}+{145\over 72}\,{\Ly^2}\nonumber \\ &&
-{2273\over 216}\,{\Ly}\Biggr ){}\,{\uutps } \nonumber \\ &&
+{}\Biggl ({11\over 6}\,{\Lu}\,{\pi^2}-{2273\over 432}\,{\Lx}-{35\over 36}\,{\pi^4}-{865\over 144}\,{\pi^2}+{35\over 12}\,{\zeta_3}-{41\over 2}\,{\Lidz}-{35\over 6}\,{\Licy}+{20}\,{\Lidx}\nonumber \\ &&
+{41\over 2}\,{\Lidy}-{35\over 12}\,{\Licx}-{10}\,{\Licx}\,{\Ly}-{7\over 4}\,{\pi^2}\,{\Ly}-{20}\,{\Licy}\,{\Ly}+{10}\,{\zeta_3}\,{\Ly}-{13\over 4}\,{\pi^2}\,{\Ly^2}\nonumber \\ &&
-{7\over 4}\,{\Lx^2}\,{\Ly}+{35\over 12}\,{\Lx}\,{\Libx}-{8}\,{\Lx}\,{\zeta_3}+{1\over 4}\,{\Lx^2}\,{\Libx}+{55\over 36}\,{\Lx}\,{\pi^2}+{11\over 6}\,{\Lu}\,{\Lx^2}-{3\over 2}\,{\Lx}\,{\Ly^3}\nonumber \\ &&
-{7\over 4}\,{\pi^2}\,{\Lx^2}+{35\over 6}\,{\Liby}\,{\Ly}-{15\over 4}\,{\pi^2}\,{\Libx}+{10}\,{\Licy}\,{\Lx}-{121\over 18}\,{\Lu}\,{\Lx}-{11\over 12}\,{\Lx^4}-{865\over 144}\,{\Lx^2}\nonumber \\ &&
+{77\over 72}\,{\Lx^3}-{1\over 2}\,{\Liby}\,{\Ly^2}+{23\over 6}\,{\Lx^3}\,{\Ly}+{865\over 72}\,{\Ly}\,{\Lx}+{14\over 3}\,{\Ly^2}\,{\Lx}-{23\over 8}\,{\Lx^2}\,{\Ly^2}+{1\over 4}\,{\pi^2}\,{\Lx}\,{\Ly}\nonumber \\ &&
-{11\over 3}\,{\Lu}\,{\Lx}\,{\Ly}+{1\over 2}\,{\Liby}\,{\Lx}\,{\Ly}\Biggr ){}\,{\uutms } \nonumber \\ &&
+{}\Biggl ({}-{242\over 9}\,{\Lu^2}+{18}\,{\Lu}\,{\zeta_3}+{253\over 36}\,{\Lu}\,{\pi^2}-{4\over 27}\,{\Lx}-{7\over 40}\,{\pi^4}+{35\over 9}\,{\pi^2}-{161\over 6}\,{\zeta_3}-{1535\over 108}\,{\Lu}\nonumber \\ &&
+{2}\,{\Lidz}+{2}\,{\Lidy}-{22\over 3}\,{\Licx}+{30377\over 432}+{2}\,{\Licx}\,{\Ly}+{440\over 9}\,{\Lu}\,{\Ly}+{155\over 18}\,{\pi^2}\,{\Ly}\nonumber \\ &&
-{6}\,{\zeta_3}\,{\Ly}+{5\over 3}\,{\pi^2}\,{\Ly^2}-{2}\,{\Lx}\,{\Licx}-{41\over 3}\,{\Lx^2}\,{\Ly}+{22\over 3}\,{\Lx}\,{\Libx}+{4}\,{\Lx}\,{\zeta_3}+{\Lx^2}\,{\Libx}\nonumber \\ &&
+{223\over 36}\,{\Lx}\,{\pi^2}+{22\over 3}\,{\Lu}\,{\Lx^2}+{2}\,{\Lx}\,{\Ly^3}-{1\over 3}\,{\pi^2}\,{\Lx^2}+{\pi^2}\,{\Libx}+{44\over 3}\,{\Lu}\,{\Ly^2}-{2}\,{\Licy}\,{\Lx}\nonumber \\ &&
-{220\over 9}\,{\Lu}\,{\Lx}-{1\over 3}\,{\Lx^4}-{145\over 9}\,{\Lx^2}+{52\over 9}\,{\Lx^3}+{5\over 3}\,{\Lx^3}\,{\Ly}+{290\over 9}\,{\Ly}\,{\Lx}+{41\over 3}\,{\Ly^2}\,{\Lx}-{3\over 2}\,{\Lx^2}\,{\Ly^2}\nonumber \\ &&
+{\pi^2}\,{\Lx}\,{\Ly}-{44\over 3}\,{\Lu}\,{\Lx}\,{\Ly}+{2}\,{\Liby}\,{\Lx}\,{\Ly}-{\Ly^4}-{82\over 9}\,{\Ly^3}-{290\over 9}\,{\Ly^2}+{8\over 27}\,{\Ly}\Biggr ){}\,{\tspst } \nonumber \\ &&
+{}\Biggl ({2995\over 432}\,{\Lx}+{38\over 45}\,{\pi^4}+{509\over 48}\,{\pi^2}+{83\over 12}\,{\zeta_3}+{20}\,{\Lidz}-{83\over 6}\,{\Licy}-{16}\,{\Lidx}-{20}\,{\Lidy}\nonumber \\ &&
-{83\over 12}\,{\Licx}+{10}\,{\Licx}\,{\Ly}+{17\over 6}\,{\pi^2}\,{\Ly}+{20}\,{\Licy}\,{\Ly}-{10}\,{\zeta_3}\,{\Ly}+{2}\,{\pi^2}\,{\Ly^2}-{2}\,{\Lx}\,{\Licx}\nonumber \\ &&
+{17\over 6}\,{\Lx^2}\,{\Ly}+{83\over 12}\,{\Lx}\,{\Libx}+{4}\,{\Lx}\,{\zeta_3}+{37\over 24}\,{\Lx}\,{\pi^2}+{2}\,{\Lx}\,{\Ly^3}+{4\over 3}\,{\pi^2}\,{\Lx^2}+{83\over 6}\,{\Liby}\,{\Ly}\nonumber \\ &&
+{4}\,{\pi^2}\,{\Libx}-{10}\,{\Licy}\,{\Lx}+{275\over 18}\,{\Lu}\,{\Lx}+{2\over 3}\,{\Lx^4}+{509\over 48}\,{\Lx^2}+{5\over 24}\,{\Lx^3}-{3}\,{\Lx^3}\,{\Ly}\nonumber \\ &&
-{509\over 24}\,{\Ly}\,{\Lx}+{49\over 12}\,{\Ly^2}\,{\Lx}+{2}\,{\Lx^2}\,{\Ly^2}\Biggr ){}\,{\tsmst } 
\end{eqnarray}

\begin{eqnarray}
{B_u}&=&{ }\Biggl ({}-{11\over 6}\,{\Lu}\,{\pi^2}+{14\over 9}\,{\Lx}+{119\over 180}\,{\pi^4}+{24}\,{\pi^2}+{173\over 6}\,{\zeta_3}-{11\over 3}\,{\Lu}-{65\over 2}\,{\Lidz}-{65\over 2}\,{\Lidy}\nonumber \\ &&
+{299\over 6}\,{\Licx}-{27}\,{\Licx}\,{\Ly}+{55\over 3}\,{\Lu}\,{\Ly}-{29\over 6}\,{\pi^2}\,{\Ly}-{11}\,{\zeta_3}\,{\Ly}+{1\over 4}\,{\pi^2}\,{\Ly^2}\nonumber \\ &&
+{18}\,{\Lx}\,{\Licx}-{197\over 6}\,{\Lx^2}\,{\Ly}-{299\over 6}\,{\Lx}\,{\Libx}+{\Lx}\,{\zeta_3}-{7\over 4}\,{\Lx^2}\,{\Libx}-{11\over 6}\,{\Lx}\,{\pi^2}\nonumber \\ &&
-{11\over 6}\,{\Lu}\,{\Lx^2}-{23\over 6}\,{\Lx}\,{\Ly^3}-{5\over 4}\,{\pi^2}\,{\Lx^2}-{7\over 4}\,{\pi^2}\,{\Libx}-{11\over 3}\,{\Lu}\,{\Ly^2}+{9}\,{\Licy}\,{\Lx}-{55\over 6}\,{\Lu}\,{\Lx}\nonumber \\ &&
-{41\over 24}\,{\Lx^4}-{311\over 36}\,{\Lx^2}+{73\over 18}\,{\Lx^3}+{37\over 6}\,{\Lx^3}\,{\Ly}-{173\over 9}\,{\Ly}\,{\Lx}+{73\over 3}\,{\Ly^2}\,{\Lx}-{55\over 8}\,{\Lx^2}\,{\Ly^2}\nonumber \\ &&
+{27\over 4}\,{\pi^2}\,{\Lx}\,{\Ly}+{11\over 3}\,{\Lu}\,{\Lx}\,{\Ly}-{7\over 2}\,{\Liby}\,{\Lx}\,{\Ly}+{139\over 18}+{23\over 12}\,{\Ly^4}-{146\over 9}\,{\Ly^3}+{173\over 9}\,{\Ly^2}\nonumber \\ &&
-{28\over 9}\,{\Ly}\Biggr ){}\,{\stuu } \nonumber \\ &&
+{}\Biggl ({164771\over 2592}-{121\over 9}\,{\Lu^2}+{10}\,{\Lu}\,{\zeta_3}-{34\over 9}\,{\Lu}\,{\pi^2}-{5941\over 432}\,{\Lx}-{4\over 3}\,{\pi^4}+{1123\over 144}\,{\pi^2}-{1529\over 36}\,{\zeta_3}\nonumber \\ &&
-{7543\over 216}\,{\Lu}+{71\over 2}\,{\Lidz}+{71\over 2}\,{\Lidy}+{7\over 4}\,{\Licx}+{242\over 9}\,{\Lu}\,{\Ly}-{347\over 36}\,{\pi^2}\,{\Ly}-{38}\,{\zeta_3}\,{\Ly}\nonumber \\ &&
-{3\over 4}\,{\pi^2}\,{\Ly^2}-{23}\,{\Lx}\,{\Licx}+{11\over 4}\,{\Lx^2}\,{\Ly}-{7\over 4}\,{\Lx}\,{\Libx}+{42}\,{\Lx}\,{\zeta_3}+{21\over 4}\,{\Lx^2}\,{\Libx}\nonumber \\ &&
-{91\over 72}\,{\Lx}\,{\pi^2}+{11\over 3}\,{\Lu}\,{\Lx^2}-{23\over 6}\,{\Lx}\,{\Ly^3}+{5\over 12}\,{\pi^2}\,{\Lx^2}+{21\over 4}\,{\pi^2}\,{\Libx}-{22\over 3}\,{\Lu}\,{\Ly^2}-{46}\,{\Licy}\,{\Lx}\nonumber \\ &&
-{121\over 9}\,{\Lu}\,{\Lx}+{43\over 24}\,{\Lx^4}+{347\over 48}\,{\Lx^2}-{179\over 72}\,{\Lx^3}-{19\over 3}\,{\Lx^3}\,{\Ly}-{1169\over 72}\,{\Ly}\,{\Lx}+{59\over 12}\,{\Ly^2}\,{\Lx}\nonumber \\ &&
+{37\over 8}\,{\Lx^2}\,{\Ly^2}+{31\over 4}\,{\pi^2}\,{\Lx}\,{\Ly}+{22\over 3}\,{\Lu}\,{\Lx}\,{\Ly}+{21\over 2}\,{\Liby}\,{\Lx}\,{\Ly}+{23\over 12}\,{\Ly^4}-{59\over 18}\,{\Ly^3}\nonumber \\ &&
+{1169\over 72}\,{\Ly^2}+{5941\over 216}\,{\Ly}\Biggr ){}\,{\uutps } \nonumber \\ &&
+{}\Biggl ({}-{11\over 3}\,{\Lu}\,{\pi^2}+{207\over 16}\,{\Lx}-{319\over 90}\,{\pi^4}+{1291\over 144}\,{\pi^2}+{121\over 12}\,{\zeta_3}-{87\over 2}\,{\Lidz}-{121\over 6}\,{\Licy}\nonumber \\ &&
+{74}\,{\Lidx}+{87\over 2}\,{\Lidy}-{121\over 12}\,{\Licx}-{26}\,{\Licx}\,{\Ly}+{163\over 12}\,{\pi^2}\,{\Ly}-{52}\,{\Licy}\,{\Ly}\nonumber \\ &&
+{26}\,{\zeta_3}\,{\Ly}-{19\over 4}\,{\pi^2}\,{\Ly^2}-{21}\,{\Lx}\,{\Licx}+{163\over 12}\,{\Lx^2}\,{\Ly}+{121\over 12}\,{\Lx}\,{\Libx}-{34}\,{\Lx}\,{\zeta_3}\nonumber \\ &&
+{23\over 4}\,{\Lx^2}\,{\Libx}+{43\over 24}\,{\Lx}\,{\pi^2}-{11\over 3}\,{\Lu}\,{\Lx^2}+{1\over 6}\,{\Lx}\,{\Ly^3}-{17\over 4}\,{\pi^2}\,{\Lx^2}+{121\over 6}\,{\Liby}\,{\Ly}\nonumber \\ &&
-{95\over 12}\,{\pi^2}\,{\Libx}+{26}\,{\Licy}\,{\Lx}-{11}\,{\Lu}\,{\Lx}-{9\over 8}\,{\Lx^4}+{1291\over 144}\,{\Lx^2}-{433\over 72}\,{\Lx^3}+{17\over 2}\,{\Liby}\,{\Ly^2}\nonumber \\ &&
+{7}\,{\Lx^3}\,{\Ly}-{1291\over 72}\,{\Ly}\,{\Lx}-{7\over 2}\,{\Ly^2}\,{\Lx}-{89\over 8}\,{\Lx^2}\,{\Ly^2}+{25\over 4}\,{\pi^2}\,{\Lx}\,{\Ly}+{22\over 3}\,{\Lu}\,{\Lx}\,{\Ly}\nonumber \\ &&
-{17\over 2}\,{\Liby}\,{\Lx}\,{\Ly}\Biggr ){}\,{\uutms } \nonumber \\ &&
+{}\Biggl ({}-{2}\,{\Lu}\,{\zeta_3}-{1\over 12}\,{\Lu}\,{\pi^2}+{40\over 3}\,{\Lx}-{883\over 360}\,{\pi^4}+{329\over 36}\,{\pi^2}+{443\over 18}\,{\zeta_3}+{1502\over 27}\,{\Lu}-{19139\over 324}\nonumber \\ &&
+{18}\,{\Lidz}+{18}\,{\Lidy}+{18}\,{\Licx}\,{\Ly}+{40\over 3}\,{\pi^2}\,{\Ly}-{18}\,{\zeta_3}\,{\Ly}+{4}\,{\pi^2}\,{\Ly^2}-{18}\,{\Lx}\,{\Licx}\nonumber \\ &&
-{40\over 3}\,{\Lx^2}\,{\Ly}+{18}\,{\Lx}\,{\zeta_3}+{9}\,{\Lx^2}\,{\Libx}+{20\over 3}\,{\Lx}\,{\pi^2}+{11\over 3}\,{\pi^2}\,{\Lx^2}+{9}\,{\pi^2}\,{\Libx}-{18}\,{\Licy}\,{\Lx}\nonumber \\ &&
+{1\over 2}\,{\Lx^4}-{517\over 36}\,{\Lx^2}+{20\over 3}\,{\Lx^3}+{\Lx^3}\,{\Ly}+{7}\,{\Ly}\,{\Lx}+{19\over 2}\,{\Lx^2}\,{\Ly^2}+{\pi^2}\,{\Lx}\,{\Ly}\nonumber \\ &&
+{18}\,{\Liby}\,{\Lx}\,{\Ly}-{7}\,{\Ly^2}-{80\over 3}\,{\Ly}\Biggr ){}\,{\tspst } \nonumber \\ &&
+{}\Biggl ({11\over 6}\,{\Lu}\,{\pi^2}-{121\over 36}\,{\Lx}-{119\over 180}\,{\pi^4}-{157\over 18}\,{\pi^2}-{71\over 6}\,{\zeta_3}-{43\over 2}\,{\Lidz}+{71\over 3}\,{\Licy}\nonumber \\ &&
+{22}\,{\Lidx}+{43\over 2}\,{\Lidy}+{71\over 6}\,{\Licx}-{9}\,{\Licx}\,{\Ly}-{39\over 2}\,{\pi^2}\,{\Ly}-{18}\,{\Licy}\,{\Ly}+{9}\,{\zeta_3}\,{\Ly}\nonumber \\ &&
-{17\over 4}\,{\pi^2}\,{\Ly^2}+{2}\,{\Lx}\,{\Licx}-{39\over 2}\,{\Lx^2}\,{\Ly}-{71\over 6}\,{\Lx}\,{\Libx}-{\Lx}\,{\zeta_3}-{9\over 4}\,{\Lx^2}\,{\Libx}\nonumber \\ &&
-{137\over 18}\,{\Lx}\,{\pi^2}+{11\over 6}\,{\Lu}\,{\Lx^2}-{5\over 2}\,{\Lx}\,{\Ly^3}-{11\over 4}\,{\pi^2}\,{\Lx^2}-{71\over 3}\,{\Liby}\,{\Ly}-{5\over 4}\,{\pi^2}\,{\Libx}\nonumber \\ &&
+{9}\,{\Licy}\,{\Lx}+{11\over 6}\,{\Lu}\,{\Lx}-{43\over 24}\,{\Lx^4}-{157\over 18}\,{\Lx^2}+{221\over 18}\,{\Lx^3}-{7\over 2}\,{\Liby}\,{\Ly^2}+{29\over 6}\,{\Lx^3}\,{\Ly}\nonumber \\ &&
+{157\over 9}\,{\Ly}\,{\Lx}+{23\over 3}\,{\Ly^2}\,{\Lx}-{13\over 8}\,{\Lx^2}\,{\Ly^2}-{37\over 12}\,{\pi^2}\,{\Lx}\,{\Ly}-{11\over 3}\,{\Lu}\,{\Lx}\,{\Ly}\nonumber \\ &&
+{7\over 2}\,{\Liby}\,{\Lx}\,{\Ly}\Biggr ){}\,{\tsmst } -{\Ly^2}\,{\sfiveot } +{}\Biggl ({}-{\Ly^2}+{2}\,{\Ly}\,{\Lx}-{\pi^2}-{\Lx^2}\Biggr ){}\,{\tfiveos } 
\end{eqnarray}

\begin{eqnarray}
{C_u}&=&{ }\Biggl ({}-{22\over 3}\,{\Lu}\,{\pi^2}+{3133\over 144}\,{\Lx}-{203\over 180}\,{\pi^4}+{401\over 48}\,{\pi^2}-{1\over 4}\,{\zeta_3}+{83\over 2}\,{\Lidz}+{23\over 8}+{83\over 2}\,{\Lidy}\nonumber \\ &&
+{13\over 4}\,{\Licx}+{41}\,{\Licx}\,{\Ly}+{44\over 3}\,{\Lu}\,{\Ly}+{41\over 9}\,{\pi^2}\,{\Ly}+{9}\,{\zeta_3}\,{\Ly}-{13\over 12}\,{\pi^2}\,{\Ly^2}-{26}\,{\Lx}\,{\Licx}\nonumber \\ &&
+{41\over 3}\,{\Lx^2}\,{\Ly}-{13\over 4}\,{\Lx}\,{\Libx}+{\Lx}\,{\zeta_3}+{21\over 4}\,{\Lx^2}\,{\Libx}-{299\over 72}\,{\Lx}\,{\pi^2}-{22\over 3}\,{\Lu}\,{\Lx^2}+{3\over 2}\,{\Lx}\,{\Ly^3}\nonumber \\ &&
+{25\over 12}\,{\pi^2}\,{\Lx^2}+{21\over 4}\,{\pi^2}\,{\Libx}-{44\over 3}\,{\Lu}\,{\Ly^2}-{11}\,{\Licy}\,{\Lx}-{22\over 3}\,{\Lu}\,{\Lx}+{55\over 24}\,{\Lx^4}+{919\over 144}\,{\Lx^2}\nonumber \\ &&
-{209\over 72}\,{\Lx^3}-{47\over 6}\,{\Lx^3}\,{\Ly}-{259\over 72}\,{\Ly}\,{\Lx}-{161\over 6}\,{\Ly^2}\,{\Lx}+{125\over 8}\,{\Lx^2}\,{\Ly^2}-{33\over 4}\,{\pi^2}\,{\Lx}\,{\Ly}\nonumber \\ &&
+{44\over 3}\,{\Lu}\,{\Lx}\,{\Ly}+{21\over 2}\,{\Liby}\,{\Lx}\,{\Ly}-{3\over 4}\,{\Ly^4}+{161\over 9}\,{\Ly^3}+{259\over 72}\,{\Ly^2}-{3133\over 72}\,{\Ly}\Biggr ){}\,{\stuu } \nonumber \\ &&
+{}\Biggl ({5}\,{\Lu}\,{\zeta_3}-{145\over 24}\,{\Lu}\,{\pi^2}+{743\over 48}\,{\Lx}-{113\over 240}\,{\pi^4}+{1631\over 144}\,{\pi^2}-{4\over 9}\,{\zeta_3}+{5297\over 216}\,{\Lu}+{19\over 2}\,{\Lidz}\nonumber \\ &&
+{19\over 2}\,{\Lidy}-{41393\over 2592}+{21\over 4}\,{\Licx}+{19}\,{\Licx}\,{\Ly}+{11}\,{\Lu}\,{\Ly}+{5\over 2}\,{\pi^2}\,{\Ly}-{\zeta_3}\,{\Ly}\nonumber \\ &&
+{5\over 12}\,{\pi^2}\,{\Ly^2}-{7}\,{\Lx}\,{\Licx}+{17\over 2}\,{\Lx^2}\,{\Ly}-{21\over 4}\,{\Lx}\,{\Libx}-{2}\,{\Lx}\,{\zeta_3}+{9\over 4}\,{\Lx^2}\,{\Libx}\nonumber \\ &&
-{27\over 8}\,{\Lx}\,{\pi^2}-{11\over 2}\,{\Lu}\,{\Lx^2}+{5\over 6}\,{\Lx}\,{\Ly^3}+{3\over 4}\,{\pi^2}\,{\Lx^2}+{9\over 4}\,{\pi^2}\,{\Libx}-{11}\,{\Lu}\,{\Ly^2}+{5}\,{\Licy}\,{\Lx}\nonumber \\ &&
-{11\over 2}\,{\Lu}\,{\Lx}+{7\over 12}\,{\Lx^4}+{97\over 48}\,{\Lx^2}-{17\over 8}\,{\Lx^3}-{4\over 3}\,{\Lx^3}\,{\Ly}-{145\over 24}\,{\Ly}\,{\Lx}-{18}\,{\Ly^2}\,{\Lx}+{57\over 8}\,{\Lx^2}\,{\Ly^2}\nonumber \\ &&
-{61\over 12}\,{\pi^2}\,{\Lx}\,{\Ly}+{11}\,{\Lu}\,{\Lx}\,{\Ly}+{9\over 2}\,{\Liby}\,{\Lx}\,{\Ly}-{5\over 12}\,{\Ly^4}+{12}\,{\Ly^3}+{145\over 24}\,{\Ly^2}\nonumber \\ &&
-{743\over 24}\,{\Ly}\Biggr ){}\,{\uutps } \nonumber \\ &&
+{}\Biggl ({}-{11\over 6}\,{\Lu}\,{\pi^2}+{67\over 16}\,{\Lx}+{14\over 45}\,{\pi^4}+{1141\over 144}\,{\pi^2}+{9\over 4}\,{\zeta_3}+{21\over 2}\,{\Lidz}-{9\over 2}\,{\Licy}-{8}\,{\Lidx}\nonumber \\ &&
-{21\over 2}\,{\Lidy}-{9\over 4}\,{\Licx}+{5}\,{\Licx}\,{\Ly}+{17\over 3}\,{\pi^2}\,{\Ly}+{10}\,{\Licy}\,{\Ly}-{5}\,{\zeta_3}\,{\Ly}+{3\over 4}\,{\pi^2}\,{\Ly^2}\nonumber \\ &&
-{3}\,{\Lx}\,{\Licx}+{17\over 3}\,{\Lx^2}\,{\Ly}+{9\over 4}\,{\Lx}\,{\Libx}+{6}\,{\Lx}\,{\zeta_3}+{7\over 4}\,{\Lx^2}\,{\Libx}+{19\over 72}\,{\Lx}\,{\pi^2}-{11\over 6}\,{\Lu}\,{\Lx^2}\nonumber \\ &&
+{3\over 2}\,{\Lx}\,{\Ly^3}+{3\over 4}\,{\pi^2}\,{\Lx^2}+{9\over 2}\,{\Liby}\,{\Ly}+{5\over 12}\,{\pi^2}\,{\Libx}-{5}\,{\Licy}\,{\Lx}+{11\over 2}\,{\Lu}\,{\Lx}+{1\over 4}\,{\Lx^4}\nonumber \\ &&
+{1141\over 144}\,{\Lx^2}-{107\over 72}\,{\Lx^3}+{1\over 2}\,{\Liby}\,{\Ly^2}-{2\over 3}\,{\Lx^3}\,{\Ly}-{1141\over 72}\,{\Ly}\,{\Lx}-{41\over 12}\,{\Ly^2}\,{\Lx}+{3\over 8}\,{\Lx^2}\,{\Ly^2}\nonumber \\ &&
-{1\over 4}\,{\pi^2}\,{\Lx}\,{\Ly}+{11\over 3}\,{\Lu}\,{\Lx}\,{\Ly}-{1\over 2}\,{\Liby}\,{\Lx}\,{\Ly}\Biggr ){}\,{\uutms } \nonumber \\ &&
+{}\Biggl ({}-{3\over 4}\,{\Lu}-{255\over 16}+{15}\,{\zeta_3}+{11\over 90}\,{\pi^4}-{35\over 12}\,{\pi^2}+{9\over 4}\,{\Lx^2}+{\Ly}\,{\Lx}-{\Ly^2}+{\Lu}\,{\pi^2}\nonumber \\ &&
-{12}\,{\Lu}\,{\zeta_3}\Biggr ){}\,{\tspst } \nonumber \\ &&
+{}\Biggl ({}-{21\over 16}\,{\Lx}-{47\over 180}\,{\pi^4}-{47\over 16}\,{\pi^2}-{11\over 4}\,{\zeta_3}-{11\over 2}\,{\Lidz}+{11\over 2}\,{\Licy}+{6}\,{\Lidx}+{11\over 2}\,{\Lidy}\nonumber \\ &&
+{11\over 4}\,{\Licx}-{3}\,{\Licx}\,{\Ly}-{3\over 2}\,{\pi^2}\,{\Ly}-{6}\,{\Licy}\,{\Ly}+{3}\,{\zeta_3}\,{\Ly}-{1\over 4}\,{\pi^2}\,{\Ly^2}-{3\over 2}\,{\Lx^2}\,{\Ly}\nonumber \\ &&
-{11\over 4}\,{\Lx}\,{\Libx}-{3}\,{\Lx}\,{\zeta_3}-{1\over 4}\,{\Lx^2}\,{\Libx}+{17\over 24}\,{\Lx}\,{\pi^2}-{1\over 2}\,{\Lx}\,{\Ly^3}-{5\over 12}\,{\pi^2}\,{\Lx^2}-{11\over 2}\,{\Liby}\,{\Ly}\nonumber \\ &&
-{7\over 12}\,{\pi^2}\,{\Libx}+{3}\,{\Licy}\,{\Lx}-{1\over 8}\,{\Lx^4}-{47\over 16}\,{\Lx^2}-{13\over 24}\,{\Lx^3}+{1\over 2}\,{\Liby}\,{\Ly^2}+{1\over 2}\,{\Lx^3}\,{\Ly}\nonumber \\ &&
+{47\over 8}\,{\Ly}\,{\Lx}-{5\over 4}\,{\Ly^2}\,{\Lx}-{5\over 8}\,{\Lx^2}\,{\Ly^2}+{1\over 4}\,{\pi^2}\,{\Lx}\,{\Ly}-{1\over 2}\,{\Liby}\,{\Lx}\,{\Ly}\Biggr ){}\,{\tsmst } \nonumber \\ &&+{\Ly^2}\,{\sfiveot } +{}\Biggl ({\Ly^2}-{2}\,{\Ly}\,{\Lx}+{\pi^2}+{\Lx^2}\Biggr ){}\,{\tfiveos } 
\end{eqnarray}

\begin{eqnarray}
{D_u}&=&{ }\Biggl ({1\over 2}+{29\over 4}\,{\Lx}+{3\over 5}\,{\pi^4}+{25\over 6}\,{\pi^2}+{15}\,{\zeta_3}-{10}\,{\Lidz}-{10}\,{\Lidy}-{19}\,{\Licx}\nonumber \\ &&
+{12}\,{\Licx}\,{\Ly}+{7\over 3}\,{\pi^2}\,{\Ly}+{12}\,{\zeta_3}\,{\Ly}+{1\over 3}\,{\pi^2}\,{\Ly^2}+{4}\,{\Lx}\,{\Licx}+{13}\,{\Lx^2}\,{\Ly}+{19}\,{\Lx}\,{\Libx}\nonumber \\ &&
-{16}\,{\Lx}\,{\zeta_3}+{\Lx^2}\,{\Libx}-{1\over 6}\,{\Lx}\,{\pi^2}+{2}\,{\Lx}\,{\Ly^3}-{2\over 3}\,{\pi^2}\,{\Lx^2}+{\pi^2}\,{\Libx}+{20}\,{\Licy}\,{\Lx}\nonumber \\ &&
-{\Lx^4}+{4}\,{\Lx^2}-{11\over 6}\,{\Lx^3}+{4}\,{\Lx^3}\,{\Ly}-{\Ly}\,{\Lx}-{9}\,{\Ly^2}\,{\Lx}+{5\over 2}\,{\Lx^2}\,{\Ly^2}-{7\over 3}\,{\pi^2}\,{\Lx}\,{\Ly}\nonumber \\ &&
+{2}\,{\Liby}\,{\Lx}\,{\Ly}-{\Ly^4}+{6}\,{\Ly^3}+{\Ly^2}-{29\over 2}\,{\Ly}\Biggr ){}\,{\stuu } \nonumber \\ &&
+{}\Biggl ({}-{6}\,{\Lu}\,{\zeta_3}+{1\over 2}\,{\Lu}\,{\pi^2}+{47\over 16}\,{\Lx}+{7\over 18}\,{\pi^4}-{3\over 16}\,{\pi^2}+{23\over 4}\,{\zeta_3}-{3\over 8}\,{\Lu}-{11\over 2}\,{\Lidz}\nonumber \\ &&
-{11\over 2}\,{\Lidy}-{13\over 4}\,{\Licx}-{187\over 32}+{7}\,{\Licx}\,{\Ly}+{7\over 6}\,{\pi^2}\,{\Ly}+{7}\,{\zeta_3}\,{\Ly}+{5\over 12}\,{\pi^2}\,{\Ly^2}\nonumber \\ &&
+{2}\,{\Lx}\,{\Licx}+{7\over 2}\,{\Lx^2}\,{\Ly}+{13\over 4}\,{\Lx}\,{\Libx}-{9}\,{\Lx}\,{\zeta_3}+{3\over 4}\,{\Lx^2}\,{\Libx}+{1\over 24}\,{\Lx}\,{\pi^2}+{7\over 6}\,{\Lx}\,{\Ly^3}\nonumber \\ &&
-{5\over 12}\,{\pi^2}\,{\Lx^2}+{3\over 4}\,{\pi^2}\,{\Libx}+{11}\,{\Licy}\,{\Lx}-{5\over 8}\,{\Lx^4}-{13\over 16}\,{\Lx^2}-{3\over 8}\,{\Lx^3}+{5\over 2}\,{\Lx^3}\,{\Ly}+{13\over 8}\,{\Ly}\,{\Lx}\nonumber \\ &&
-{5}\,{\Ly^2}\,{\Lx}+{11\over 8}\,{\Lx^2}\,{\Ly^2}-{13\over 12}\,{\pi^2}\,{\Lx}\,{\Ly}+{3\over 2}\,{\Liby}\,{\Lx}\,{\Ly}-{7\over 12}\,{\Ly^4}+{10\over 3}\,{\Ly^3}-{13\over 8}\,{\Ly^2}\nonumber \\ &&
-{47\over 8}\,{\Ly}\Biggr ){}\,{\uutps } \nonumber \\ &&
+{}\Biggl ({}-{31\over 16}\,{\Lx}+{103\over 180}\,{\pi^4}-{23\over 16}\,{\pi^2}+{19\over 4}\,{\zeta_3}+{11\over 2}\,{\Lidz}-{19\over 2}\,{\Licy}-{14}\,{\Lidx}\nonumber \\ &&
-{11\over 2}\,{\Lidy}-{19\over 4}\,{\Licx}+{3}\,{\Licx}\,{\Ly}+{2}\,{\pi^2}\,{\Ly}+{6}\,{\Licy}\,{\Ly}-{3}\,{\zeta_3}\,{\Ly}+{1\over 4}\,{\pi^2}\,{\Ly^2}\nonumber \\ &&
+{4}\,{\Lx}\,{\Licx}+{2}\,{\Lx^2}\,{\Ly}+{19\over 4}\,{\Lx}\,{\Libx}+{7}\,{\Lx}\,{\zeta_3}+{1\over 4}\,{\Lx^2}\,{\Libx}+{7\over 24}\,{\Lx}\,{\pi^2}+{1\over 2}\,{\Lx}\,{\Ly^3}\nonumber \\ &&
+{3\over 4}\,{\pi^2}\,{\Lx^2}+{19\over 2}\,{\Liby}\,{\Ly}-{3\over 4}\,{\pi^2}\,{\Libx}-{3}\,{\Licy}\,{\Lx}-{1\over 24}\,{\Lx^4}-{23\over 16}\,{\Lx^2}+{3\over 8}\,{\Lx^3}\nonumber \\ &&
-{1\over 2}\,{\Liby}\,{\Ly^2}+{1\over 6}\,{\Lx^3}\,{\Ly}+{23\over 8}\,{\Ly}\,{\Lx}+{11\over 4}\,{\Ly^2}\,{\Lx}+{5\over 8}\,{\Lx^2}\,{\Ly^2}-{19\over 12}\,{\pi^2}\,{\Lx}\,{\Ly}\nonumber \\ &&
+{1\over 2}\,{\Liby}\,{\Lx}\,{\Ly}\Biggr ){}\,{\uutms } +{}\Biggl ({2}\,{\Lx^2}-{\Ly}\,{\Lx}+{1\over 2}\,{\pi^2}+{\Ly^2}\Biggr ){}\,{\tspst } \nonumber \\ &&
+{}\Biggl ({3\over 2}\,{\Lx^2}-{3}\,{\Ly}\,{\Lx}+{3\over 2}\,{\pi^2}-{3}\,{\Lx}\Biggr ){}\,{\tsmst } -{\Ly^2}\,{\sfiveot } \nonumber \\ &&+{}\Biggl ({}-{\Ly^2}+{2}\,{\Ly}\,{\Lx}-{\pi^2}-{\Lx^2}\Biggr ){}\,{\tfiveos } 
\end{eqnarray}

\begin{eqnarray}
{E_u}&=&{ }\Biggl ({91\over 18}\,{\Lx}-{11\over 180}\,{\pi^4}-{25\over 18}\,{\pi^2}-{4}\,{\zeta_3}+{2\over 3}\,{\Lu}-{4}\,{\Licx}-{2}\,{\Licx}\,{\Ly}+{10\over 3}\,{\Lu}\,{\Ly}\nonumber \\ &&
+{2\over 3}\,{\pi^2}\,{\Ly}-{2}\,{\zeta_3}\,{\Ly}+{1\over 6}\,{\pi^2}\,{\Ly^2}-{37\over 18}+{2}\,{\Lx^2}\,{\Ly}+{4}\,{\Lx}\,{\Libx}+{2}\,{\Lx}\,{\zeta_3}+{2\over 3}\,{\Lx}\,{\pi^2}\nonumber \\ &&
-{1\over 3}\,{\Lx}\,{\Ly^3}-{2}\,{\Licy}\,{\Lx}-{5\over 3}\,{\Lu}\,{\Lx}-{2}\,{\Lx^2}+{4}\,{\Ly}\,{\Lx}-{2}\,{\Ly^2}\,{\Lx}-{1\over 2}\,{\Lx^2}\,{\Ly^2}+{2\over 3}\,{\pi^2}\,{\Lx}\,{\Ly}\nonumber \\ &&
+{1\over 6}\,{\Ly^4}+{4\over 3}\,{\Ly^3}-{4}\,{\Ly^2}-{91\over 9}\,{\Ly}\Biggr ){}\,{\stuu } \nonumber \\ &&
+{}\Biggl ({}-{44\over 9}\,{\Lu^2}+{11\over 36}\,{\Lu}\,{\pi^2}+{5\over 9}\,{\Lx}+{11\over 90}\,{\pi^4}+{133\over 108}\,{\pi^2}-{13\over 6}\,{\zeta_3}+{185\over 54}\,{\Lu}-{2\over 3}\,{\Licx}\nonumber \\ &&
+{4}\,{\Licx}\,{\Ly}+{44\over 9}\,{\Lu}\,{\Ly}+{13\over 36}\,{\pi^2}\,{\Ly}+{4}\,{\zeta_3}\,{\Ly}-{1\over 3}\,{\pi^2}\,{\Ly^2}-{1\over 4}\,{\Lx^2}\,{\Ly}+{2\over 3}\,{\Lx}\,{\Libx}\nonumber \\ &&
-{4}\,{\Lx}\,{\zeta_3}+{5\over 18}\,{\Lx}\,{\pi^2}+{1\over 3}\,{\Lu}\,{\Lx^2}+{2\over 3}\,{\Lx}\,{\Ly^3}+{2\over 3}\,{\Lu}\,{\Ly^2}+{4}\,{\Licy}\,{\Lx}-{22\over 9}\,{\Lu}\,{\Lx}\nonumber \\ &&
-{7\over 36}\,{\Lx^2}+{7\over 36}\,{\Lx^3}+{7\over 18}\,{\Ly}\,{\Lx}+{1\over 4}\,{\Ly^2}\,{\Lx}+{\Lx^2}\,{\Ly^2}-{4\over 3}\,{\pi^2}\,{\Lx}\,{\Ly}-{2\over 3}\,{\Lu}\,{\Lx}\,{\Ly}\nonumber \\ &&
-{1\over 3}\,{\Ly^4}-{1\over 6}\,{\Ly^3}-{7\over 18}\,{\Ly^2}+{1307\over 216}-{10\over 9}\,{\Ly}\Biggr ){}\,{\uutps } \nonumber \\ &&
+{}\Biggl ({}-{1\over 3}\,{\Lu}\,{\pi^2}-{5\over 9}\,{\Lx}+{11\over 30}\,{\pi^4}+{79\over 36}\,{\pi^2}-{2\over 3}\,{\zeta_3}+{8}\,{\Lidz}+{4\over 3}\,{\Licy}-{8}\,{\Lidx}\nonumber \\ &&
-{8}\,{\Lidy}+{2\over 3}\,{\Licx}+{4}\,{\Licx}\,{\Ly}+{1\over 4}\,{\pi^2}\,{\Ly}+{8}\,{\Licy}\,{\Ly}-{4}\,{\zeta_3}\,{\Ly}+{\pi^2}\,{\Ly^2}\nonumber \\ &&
+{1\over 4}\,{\Lx^2}\,{\Ly}-{2\over 3}\,{\Lx}\,{\Libx}+{4}\,{\Lx}\,{\zeta_3}-{5\over 18}\,{\Lx}\,{\pi^2}-{1\over 3}\,{\Lu}\,{\Lx^2}+{2\over 3}\,{\Lx}\,{\Ly^3}+{2\over 3}\,{\pi^2}\,{\Lx^2}\nonumber \\ &&
-{4\over 3}\,{\Liby}\,{\Ly}+{4\over 3}\,{\pi^2}\,{\Libx}-{4}\,{\Licy}\,{\Lx}+{22\over 9}\,{\Lu}\,{\Lx}+{1\over 3}\,{\Lx^4}+{79\over 36}\,{\Lx^2}-{7\over 36}\,{\Lx^3}\nonumber \\ &&
-{4\over 3}\,{\Lx^3}\,{\Ly}-{79\over 18}\,{\Ly}\,{\Lx}-{11\over 12}\,{\Ly^2}\,{\Lx}+{\Lx^2}\,{\Ly^2}+{2\over 3}\,{\Lu}\,{\Lx}\,{\Ly}\Biggr ){}\,{\uutms } \nonumber \\ &&
+{}\Biggl ({88\over 9}\,{\Lu^2}-{28\over 9}\,{\Lx}-{221\over 27}\,{\Lu}+{4\over 3}\,{\Licx}+{13\over 3}\,{\zeta_3}-{67\over 54}\,{\pi^2}-{124\over 9}\,{\Lu}\,{\Ly}-{17\over 18}\,{\Lx}\,{\pi^2}\nonumber \\ &&
+{5\over 3}\,{\Lx^2}\,{\Ly}-{4\over 3}\,{\Lx}\,{\Libx}-{4\over 3}\,{\Lu}\,{\Lx^2}-{8\over 3}\,{\Lu}\,{\Ly^2}+{62\over 9}\,{\Lu}\,{\Lx}+{71\over 18}\,{\Lx^2}-{7\over 9}\,{\Lx^3}-{71\over 9}\,{\Ly}\,{\Lx}\nonumber \\ &&
-{5\over 3}\,{\Ly^2}\,{\Lx}+{8\over 3}\,{\Lu}\,{\Lx}\,{\Ly}+{10\over 9}\,{\Ly^3}+{71\over 9}\,{\Ly^2}-{10\over 9}\,{\pi^2}\,{\Ly}-{863\over 108}-{23\over 18}\,{\Lu}\,{\pi^2}\nonumber \\ &&
+{56\over 9}\,{\Ly}\Biggr ){}\,{\tspst } \nonumber \\ &&+{}\Biggl ({37\over 18}\,{\Lx}-{11\over 60}\,{\pi^4}-{23\over 6}\,{\pi^2}-{8\over 3}\,{\zeta_3}-{4}\,{\Lidz}+{16\over 3}\,{\Licy}+{4}\,{\Lidx}+{4}\,{\Lidy}\nonumber \\ &&
+{8\over 3}\,{\Licx}-{2}\,{\Licx}\,{\Ly}-{4\over 3}\,{\pi^2}\,{\Ly}-{4}\,{\Licy}\,{\Ly}+{2}\,{\zeta_3}\,{\Ly}-{1\over 2}\,{\pi^2}\,{\Ly^2}-{4\over 3}\,{\Lx^2}\,{\Ly}\nonumber \\ &&
-{8\over 3}\,{\Lx}\,{\Libx}-{2}\,{\Lx}\,{\zeta_3}-{1\over 2}\,{\Lx}\,{\pi^2}-{1\over 3}\,{\Lx}\,{\Ly^3}-{1\over 3}\,{\pi^2}\,{\Lx^2}-{16\over 3}\,{\Liby}\,{\Ly}-{2\over 3}\,{\pi^2}\,{\Libx}\nonumber \\ &&
+{2}\,{\Licy}\,{\Lx}-{47\over 9}\,{\Lu}\,{\Lx}-{1\over 6}\,{\Lx^4}-{23\over 6}\,{\Lx^2}+{2\over 3}\,{\Lx^3}\,{\Ly}+{23\over 3}\,{\Ly}\,{\Lx}-{4\over 3}\,{\Ly^2}\,{\Lx}\nonumber \\ &&
-{1\over 2}\,{\Lx^2}\,{\Ly^2}\Biggr ){}\,{\tsmst } 
\end{eqnarray}

\begin{eqnarray}
{F_u}&=&{ }\Biggl ({1\over 3}\,{\Lu}\,{\pi^2}-{50\over 9}\,{\Lx}-{11\over 45}\,{\pi^4}+{41\over 9}\,{\pi^2}+{52\over 3}\,{\zeta_3}+{2\over 3}\,{\Lu}-{19\over 18}+{52\over 3}\,{\Licx}-{8}\,{\Licx}\,{\Ly}\nonumber \\ &&
-{10\over 3}\,{\Lu}\,{\Ly}-{85\over 36}\,{\pi^2}\,{\Ly}-{8}\,{\zeta_3}\,{\Ly}+{2\over 3}\,{\pi^2}\,{\Ly^2}-{83\over 12}\,{\Lx^2}\,{\Ly}-{52\over 3}\,{\Lx}\,{\Libx}+{8}\,{\Lx}\,{\zeta_3}\nonumber \\ &&
-{115\over 36}\,{\Lx}\,{\pi^2}+{1\over 3}\,{\Lu}\,{\Lx^2}-{4\over 3}\,{\Lx}\,{\Ly^3}+{2\over 3}\,{\Lu}\,{\Ly^2}-{8}\,{\Licy}\,{\Lx}+{5\over 3}\,{\Lu}\,{\Lx}+{49\over 18}\,{\Lx^2}\nonumber \\ &&
-{11\over 36}\,{\Lx^3}-{43\over 9}\,{\Ly}\,{\Lx}+{21\over 4}\,{\Ly^2}\,{\Lx}-{2}\,{\Lx^2}\,{\Ly^2}+{8\over 3}\,{\pi^2}\,{\Lx}\,{\Ly}-{2\over 3}\,{\Lu}\,{\Lx}\,{\Ly}+{2\over 3}\,{\Ly^4}-{7\over 2}\,{\Ly^3}\nonumber \\ &&
+{43\over 9}\,{\Ly^2}+{100\over 9}\,{\Ly}\Biggr ){}\,{\stuu } \nonumber \\ &&
+{}\Biggl ({44\over 9}\,{\Lu^2}-{3661\over 324}+{7\over 9}\,{\Lu}\,{\pi^2}-{115\over 36}\,{\Lx}-{11\over 45}\,{\pi^4}-{367\over 108}\,{\pi^2}+{14\over 9}\,{\zeta_3}+{7\over 9}\,{\Lu}-{8\over 3}\,{\Licx}\nonumber \\ &&
-{8}\,{\Licx}\,{\Ly}-{88\over 9}\,{\Lu}\,{\Ly}-{1\over 12}\,{\pi^2}\,{\Ly}-{8}\,{\zeta_3}\,{\Ly}+{2\over 3}\,{\pi^2}\,{\Ly^2}+{1\over 4}\,{\Lx^2}\,{\Ly}+{8\over 3}\,{\Lx}\,{\Libx}\nonumber \\ &&
+{8}\,{\Lx}\,{\zeta_3}-{2\over 3}\,{\Lu}\,{\Lx^2}-{4\over 3}\,{\Lx}\,{\Ly^3}+{4\over 3}\,{\Lu}\,{\Ly^2}-{8}\,{\Licy}\,{\Lx}+{44\over 9}\,{\Lu}\,{\Lx}-{11\over 9}\,{\Lx^2}-{1\over 18}\,{\Lx^3}\nonumber \\ &&
+{53\over 18}\,{\Ly}\,{\Lx}+{9\over 4}\,{\Ly^2}\,{\Lx}-{2}\,{\Lx^2}\,{\Ly^2}+{8\over 3}\,{\pi^2}\,{\Lx}\,{\Ly}-{4\over 3}\,{\Lu}\,{\Lx}\,{\Ly}+{2\over 3}\,{\Ly^4}-{3\over 2}\,{\Ly^3}-{53\over 18}\,{\Ly^2}\nonumber \\ &&
+{115\over 18}\,{\Ly}\Biggr ){}\,{\uutps } \nonumber \\ &&
+{}\Biggl ({2\over 3}\,{\Lu}\,{\pi^2}-{23\over 4}\,{\Lx}-{11\over 15}\,{\pi^4}-{19\over 6}\,{\pi^2}+{4\over 3}\,{\zeta_3}-{16}\,{\Lidz}-{8\over 3}\,{\Licy}+{16}\,{\Lidx}\nonumber \\ &&
+{16}\,{\Lidy}-{4\over 3}\,{\Licx}-{8}\,{\Licx}\,{\Ly}-{19\over 12}\,{\pi^2}\,{\Ly}-{16}\,{\Licy}\,{\Ly}+{8}\,{\zeta_3}\,{\Ly}-{2}\,{\pi^2}\,{\Ly^2}\nonumber \\ &&
-{19\over 12}\,{\Lx^2}\,{\Ly}+{4\over 3}\,{\Lx}\,{\Libx}-{8}\,{\Lx}\,{\zeta_3}-{1\over 18}\,{\Lx}\,{\pi^2}+{2\over 3}\,{\Lu}\,{\Lx^2}-{4\over 3}\,{\Lx}\,{\Ly^3}-{4\over 3}\,{\pi^2}\,{\Lx^2}\nonumber \\ &&
+{8\over 3}\,{\Liby}\,{\Ly}-{8\over 3}\,{\pi^2}\,{\Libx}+{8}\,{\Licy}\,{\Lx}+{2}\,{\Lu}\,{\Lx}-{2\over 3}\,{\Lx^4}-{19\over 6}\,{\Lx^2}+{13\over 18}\,{\Lx^3}+{8\over 3}\,{\Lx^3}\,{\Ly}\nonumber \\ &&
+{19\over 3}\,{\Ly}\,{\Lx}+{35\over 12}\,{\Ly^2}\,{\Lx}-{2}\,{\Lx^2}\,{\Ly^2}-{4\over 3}\,{\Lu}\,{\Lx}\,{\Ly}\Biggr ){}\,{\uutms } \nonumber \\ &&
+{}\Biggl ({}-{4\over 3}\,{\Lx}-{263\over 27}\,{\Lu}-{1\over 9}\,{\zeta_3}-{1\over 3}\,{\pi^2}-{2\over 3}\,{\Lx}\,{\pi^2}+{4\over 3}\,{\Lx^2}\,{\Ly}+{85\over 18}\,{\Lx^2}-{2\over 3}\,{\Lx^3}-{3}\,{\Ly}\,{\Lx}\nonumber \\ &&
+{3}\,{\Ly^2}-{4\over 3}\,{\pi^2}\,{\Ly}+{4085\over 324}-{1\over 6}\,{\Lu}\,{\pi^2}+{8\over 3}\,{\Ly}\Biggr ){}\,{\tspst } \nonumber \\ &&
+{}\Biggl ({19\over 9}\,{\Lx}+{8\over 3}\,{\Licy}+{4\over 3}\,{\Licx}-{4\over 3}\,{\zeta_3}+{7}\,{\pi^2}+{5\over 12}\,{\Lx}\,{\pi^2}+{9\over 4}\,{\Lx^2}\,{\Ly}-{4\over 3}\,{\Lx}\,{\Libx}\nonumber \\ &&
-{1\over 3}\,{\Lu}\,{\Lx^2}-{8\over 3}\,{\Liby}\,{\Ly}-{1\over 3}\,{\Lu}\,{\Lx}+{7}\,{\Lx^2}-{49\over 36}\,{\Lx^3}-{14}\,{\Ly}\,{\Lx}-{43\over 12}\,{\Ly^2}\,{\Lx}\nonumber \\ &&
+{2\over 3}\,{\Lu}\,{\Lx}\,{\Ly}+{9\over 4}\,{\pi^2}\,{\Ly}-{1\over 3}\,{\Lu}\,{\pi^2}\Biggr ){}\,{\tsmst } -{3}\,{\Ly^2}\,{\sfiveot } \nonumber \\ &&
+{}\Biggl ({}-{3}\,{\pi^2}-{3}\,{\Lx^2}-{3}\,{\Ly^2}+{6}\,{\Ly}\,{\Lx}\Biggr ){}\,{\tfiveos } 
\end{eqnarray}

\begin{eqnarray}
{G_u}&=&{ }\Biggl ({4\over 3}\,{\Lu}\,{\pi^2}+{71\over 9}\,{\Lx}+{11\over 15}\,{\pi^4}-{20}\,{\pi^2}-{52}\,{\zeta_3}-{52}\,{\Licx}+{24}\,{\Licx}\,{\Ly}-{8\over 3}\,{\Lu}\,{\Ly}\nonumber \\ &&
+{64\over 9}\,{\pi^2}\,{\Ly}+{24}\,{\zeta_3}\,{\Ly}-{2}\,{\pi^2}\,{\Ly^2}+{64\over 3}\,{\Lx^2}\,{\Ly}+{52}\,{\Lx}\,{\Libx}-{24}\,{\Lx}\,{\zeta_3}+{85\over 9}\,{\Lx}\,{\pi^2}\nonumber \\ &&
+{4\over 3}\,{\Lu}\,{\Lx^2}+{4}\,{\Lx}\,{\Ly^3}+{8\over 3}\,{\Lu}\,{\Ly^2}+{24}\,{\Licy}\,{\Lx}+{4\over 3}\,{\Lu}\,{\Lx}-{37\over 9}\,{\Lx^2}+{7\over 9}\,{\Lx^3}+{86\over 9}\,{\Ly}\,{\Lx}\nonumber \\ &&
-{50\over 3}\,{\Ly^2}\,{\Lx}+{6}\,{\Lx^2}\,{\Ly^2}-{8}\,{\pi^2}\,{\Lx}\,{\Ly}-{8\over 3}\,{\Lu}\,{\Lx}\,{\Ly}-{2}\,{\Ly^4}+{100\over 9}\,{\Ly^3}-{86\over 9}\,{\Ly^2}-{142\over 9}\,{\Ly}\Biggr ){}\,{\stuu } \nonumber \\ &&
+{}\Biggl ({11\over 12}\,{\Lu}\,{\pi^2}+{79\over 12}\,{\Lx}+{11\over 15}\,{\pi^4}+{151\over 36}\,{\pi^2}+{71\over 18}\,{\zeta_3}+{3401\over 648}-{227\over 54}\,{\Lu}+{4}\,{\Licx}\nonumber \\ &&
+{24}\,{\Licx}\,{\Ly}-{2}\,{\Lu}\,{\Ly}+{5\over 3}\,{\pi^2}\,{\Ly}+{24}\,{\zeta_3}\,{\Ly}-{2}\,{\pi^2}\,{\Ly^2}-{2}\,{\Lx^2}\,{\Ly}-{4}\,{\Lx}\,{\Libx}\nonumber \\ &&
-{24}\,{\Lx}\,{\zeta_3}-{1\over 12}\,{\Lx}\,{\pi^2}+{\Lu}\,{\Lx^2}+{4}\,{\Lx}\,{\Ly^3}+{2}\,{\Lu}\,{\Ly^2}+{24}\,{\Licy}\,{\Lx}+{\Lu}\,{\Lx}+{31\over 6}\,{\Lx^2}\nonumber \\ &&
+{7\over 12}\,{\Lx^3}-{65\over 6}\,{\Ly}\,{\Lx}-{3\over 2}\,{\Ly^2}\,{\Lx}+{6}\,{\Lx^2}\,{\Ly^2}-{8}\,{\pi^2}\,{\Lx}\,{\Ly}-{2}\,{\Lu}\,{\Lx}\,{\Ly}-{2}\,{\Ly^4}+{\Ly^3}\nonumber \\ &&
+{65\over 6}\,{\Ly^2}-{79\over 6}\,{\Ly}\Biggr ){}\,{\uutps } \nonumber \\ &&
+{}\Biggl ({1\over 3}\,{\Lu}\,{\pi^2}+{41\over 4}\,{\Lx}+{11\over 5}\,{\pi^4}+{85\over 18}\,{\pi^2}-{4}\,{\zeta_3}+{48}\,{\Lidz}+{8}\,{\Licy}-{48}\,{\Lidx}\nonumber \\ &&
-{48}\,{\Lidy}+{4}\,{\Licx}+{24}\,{\Licx}\,{\Ly}-{2\over 3}\,{\pi^2}\,{\Ly}+{48}\,{\Licy}\,{\Ly}-{24}\,{\zeta_3}\,{\Ly}+{6}\,{\pi^2}\,{\Ly^2}\nonumber \\ &&
-{2\over 3}\,{\Lx^2}\,{\Ly}-{4}\,{\Lx}\,{\Libx}+{24}\,{\Lx}\,{\zeta_3}-{17\over 36}\,{\Lx}\,{\pi^2}+{1\over 3}\,{\Lu}\,{\Lx^2}+{4}\,{\Lx}\,{\Ly^3}+{4}\,{\pi^2}\,{\Lx^2}\nonumber \\ &&
-{8}\,{\Liby}\,{\Ly}+{8}\,{\pi^2}\,{\Libx}-{24}\,{\Licy}\,{\Lx}-{\Lu}\,{\Lx}+{2}\,{\Lx^4}+{85\over 18}\,{\Lx^2}+{7\over 36}\,{\Lx^3}-{8}\,{\Lx^3}\,{\Ly}\nonumber \\ &&
-{85\over 9}\,{\Ly}\,{\Lx}-{10\over 3}\,{\Ly^2}\,{\Lx}+{6}\,{\Lx^2}\,{\Ly^2}-{2\over 3}\,{\Lu}\,{\Lx}\,{\Ly}\Biggr ){}\,{\uutms } \nonumber \\ &&
+{}\Biggl ({}-{9\over 2}\,{\Lx^2}+{9}\,{\Ly}\,{\Lx}-{9\over 2}\,{\pi^2}-{9}\,{\Ly^2}\Biggr ){}\,{\tspst } \nonumber \\ &&
+{}\Biggl ({4}\,{\Lx}\,{\Libx}-{4}\,{\Licx}-{8}\,{\Licy}+{4}\,{\zeta_3}-{27\over 2}\,{\Lx^2}+{27}\,{\Ly}\,{\Lx}+{4}\,{\Ly^2}\,{\Lx}+{2\over 3}\,{\Lx}\,{\pi^2}\nonumber \\ &&
+{8}\,{\Liby}\,{\Ly}-{27\over 2}\,{\pi^2}-{8}\,{\Lx}\Biggr ){}\,{\tsmst } +{9}\,{\Ly^2}\,{\sfiveot } \nonumber \\ &&+{}\Biggl ({9}\,{\Lx^2}+{9}\,{\pi^2}-{18}\,{\Ly}\,{\Lx}+{9}\,{\Ly^2}\Biggr ){}\,{\tfiveos } 
\end{eqnarray}

\begin{eqnarray}
{H_u}&=&{ }\Biggl ({1\over 12}\,{\Lx^2}-{5\over 27}\,{\Lx}+{1\over 6}\,{\Ly^2}-{1\over 6}\,{\Ly}\,{\Lx}-{4\over 9}\,{\Lu}\,{\Ly}-{20\over 27}\,{\Lu}-{7\over 108}\,{\pi^2}+{2\over 9}\,{\Lu}\,{\Lx}+{4\over 9}\,{\Lu^2}\nonumber \\ &&
+{10\over 27}\,{\Ly}\Biggr ){}\,{\uutps } \nonumber \\ &&+{}\Biggl ({}-{1\over 12}\,{\pi^2}-{1\over 12}\,{\Lx^2}+{5\over 27}\,{\Lx}+{1\over 6}\,{\Ly}\,{\Lx}-{2\over 9}\,{\Lu}\,{\Lx}\Biggr ){}\,{\uutms } \nonumber \\ &&+{}\Biggl ({}-{8\over 9}\,{\Lu^2}+{40\over 27}\,{\Lu}-{1\over 6}\,{\Lx^2}-{20\over 27}\,{\Ly}+{1\over 3}\,{\Ly}\,{\Lx}+{7\over 54}\,{\pi^2}-{1\over 3}\,{\Ly^2}+{10\over 27}\,{\Lx}\nonumber \\ &&
+{8\over 9}\,{\Lu}\,{\Ly}-{4\over 9}\,{\Lu}\,{\Lx}\Biggr ){}\,{\tspst } \nonumber \\ &&+{}\Biggl ({1\over 6}\,{\Lx^2}-{1\over 3}\,{\Ly}\,{\Lx}+{1\over 6}\,{\pi^2}-{10\over 27}\,{\Lx}+{4\over 9}\,{\Lu}\,{\Lx}\Biggr ){}\,{\tsmst } 
\end{eqnarray}

\begin{eqnarray}
{I_u}&=&{ }\Biggl ({8\over 9}\,{\Lu}\,{\Ly}+{1\over 3}\,{\Ly}\,{\Lx}-{4\over 9}\,{\Lu}\,{\Lx}+{10\over 27}\,{\Lx}+{5\over 54}\,{\pi^2}-{1\over 9}\,{\Lx^2}+{20\over 27}\,{\Lu}-{20\over 27}\,{\Ly}\nonumber \\ &&
-{1\over 3}\,{\Ly^2}-{4\over 9}\,{\Lu^2}\Biggr ){}\,{\uutps } +{}\Biggl ({1\over 18}\,{\pi^2}+{1\over 18}\,{\Lx^2}-{1\over 9}\,{\Ly}\,{\Lx}\Biggr ){}\,{\uutms } \nonumber \\ &&
-{1\over 9}\,{}\Biggl ({\Lx}^2-{\pi}^2\Biggr ){}\,{\tspst } 
+{}\Biggl ({2\over 9}\,{\Ly}\,{\Lx}-{1\over 9}\,{\pi^2}-{1\over 9}\,{\Lx^2}\Biggr ){}\,{\tsmst } 
\end{eqnarray}

\section{One-loop self-interference}
\label{sec:one}
In this section, we present the self-interference of the one-loop 
amplitude of Eq.~(\ref{eq:self_interf}) in terms of the one-loop box in
$D=6-2\epsilon$ 
dimensions  
and the one-loop bubble in $D=4-2\epsilon$ dimensions. 
The $\ep$ expansion of these
integrals is given in the Appendix and can be inserted directly into the
following expressions.
We can write
\begin{equation}
\C^{8 \, (1 \times 1)}(s,t,u) = -\PSQUARE\,  \Ttwo(s,t,u)\,s^2tu +
2 \Re \left\{ \PLOOP \right\}+\FINITE,
\end{equation}
where the self-interference of the singular terms of the amplitude is 
\begin{eqnarray}
\label{eq:PSQUARE}
\PSQUARE &=& \frac{VN}{4} \absq{\PC(s, t, u)} \nonumber \\
&&+\frac{V}{4} 2 \Re\Big\{ \PC(s, t, u)^\dag \left[ \PA(s, t, u)+\RA(s, t,u) 
+ (t \leftrightarrow u)\right]\Big\}  \nonumber \\
&&+\frac{V^2}{4N} \left\{ \absq{\PA(s, t, u)+\RA(s, t, u)} 
+(t \leftrightarrow u) \right\}  
\nonumber \\
&& 
-\frac{V}{4N} \, 2\Re\left\{ \lq \PA(s, t, u)^\dag+\RA(s, t, u)^\dag \rq \lq \PA(s, u, t)+\RA(s, u, t)  \rq \right\},\nonumber\\
\end{eqnarray}
and the interference of the singular terms with the one-loop amplitude is 
\begin{eqnarray}
\label{eq:PLOOP}
\PLOOP &=& -\frac{s}{2}\left\{ 
\frac{VN}{4} \PC(s, t, u)^\dag u \Lthree(s, t, u) \right.
\nonumber \\
&& 
+\frac{V}{4} \lq 
\left(\PA(s, t, u)^\dag +\RA(s, t, u)^\dag \right) u \Lthree(s, t, u)
\right. \nonumber \\
&&\left. +
\PC(s, t, u)^\dag u \Ltwo(s, t, u)+
(t \leftrightarrow u)   
\rq
\nonumber \\
&& 
+\frac{V^2}{4N} \lq
\left( \PA(s, t, u)^\dag + \RA(s, t, u)^\dag \right) u \Ltwo(s, t, u)
+(t \leftrightarrow u)   
\rq
\nonumber \\
&& 
\left.
-\frac{V}{4N} \lq
\left( \PA(s, u, t)^\dag + \RA(s, u, t)^\dag \right) u \Ltwo(s, t, u)
+(t \leftrightarrow u)   
\rq \right\},
\end{eqnarray}
where the functions $\Ltwo$ and $\Lthree$ are defined in
Eqs.~(\ref{eq:Ltwodef}) and~(\ref{eq:Lthreedef}) respectively. In
Eqs.~(\ref{eq:PSQUARE}) and~(\ref{eq:PLOOP})
\begin{eqnarray}
\label{eq:pa}
\PA(s, t, u) &=& \frac{1-2\ep}{2st}
\left\{
\bnot \lq
\bubs+\bubt
\rq
+\lq \frac{V}{N}\left( \frac{1}{\ep}+ \frac{3}{2}\right) 
-\frac{3N}{2}  \right]
\bubs \right.
\nonumber \\
&& 
\left. +  N \left( \frac{2}{\ep} +\frac{3}{2} \right)  \bubt
\right\}, \\
\label{eq:pc}
\PC(s, t, u)&=&-\frac{1-2\ep}{\ep} 
\lq
\frac{1}{tu} \bubs
+\frac{1}{su} \bubt
+\frac{1}{st} \bubu
\rq 
\end{eqnarray}
are infrared divergent while
\begin{eqnarray}
\RA(s, t, u)=-\frac{1}{2st}\bnot \left\{ 
\frac{(2-\ep)(3-8\ep+8\ep^2)}{(1-\ep)(3-2\ep)}\bubs-\frac{2}{\ep}
\right\}, 
\end{eqnarray}
originates from ultraviolet terms which are rendered finite 
after renormalisation in the \MSbar\ scheme.

Finally, the finite contribution is given by
\begin{eqnarray}
\label{eq:finiteone}
\FINITE &=&  
\frac{VN}{4}\, \MCC(s, t, u)+ \frac{V}{2} \, 2\Re \left\{ \MCA(s, t, u) \right\}  + \frac{V^2}{4N} \MAA(s, t, u) -\frac{V}{4N} \MAB(s, t, u) 
\nonumber \\
&& +\left( t \leftrightarrow u\right),
\end{eqnarray}
where
\begin{eqnarray}
\label{eq:FCC}
\MCC(s, t, u) &=& 
\lq 
\frac{2 u \left( 8 t^4 +4 s^4 +12 s t^3 +13 s^2 t^2 +4 s^3 t\right)}{s^2 t} 
\rq 
\absq{\bst}
\nonumber \\
&& + \lq
\frac{4 u \left( 2 s^2 + t s + 4 t^2 \right)}{t}
\rq 
2 \Re \left\{ \bst^\dag \btu\right\}
\nonumber \\
&& + \lq
\frac{4 t u \left( 5 s^2 +4 t s +4 t^2 \right)}{s^2}
\rq 
\bst^\dag \bsu 
\nonumber \\
&& +\lq 
\frac{4 s^2 \left( t^2+u^2 \right)}{t u}
\rq \absq{ \btu}, 
\end{eqnarray}

\begin{eqnarray}
\MCA(s, t, u) &=& \bnot \fbub{t,s}^\dag
\left\{ 
\lq  \frac{2 \left(t^2+u^2 \right)}{t} \rq \btu +
\lq  \frac{u^2 \left( 2 s^2 +t s + 4 t^2 \right)}{ s^2 t }  \rq \bst 
\right. \nonumber \\
&& \hspace{-2.0cm} \left.
+ \lq \frac{t \left( 4 t^2 +5 s^2 +7 t s\right)}{s^2} \rq \bsu \right\} 
\nonumber \\
&& \hspace{-2.0cm} 
+\lq
\frac{(u-t) ( 2 s^4+3 s^3 t -7 su t^2  +4 t^4) N} {s^2 t}
+\frac{2 s^4 +2 t^4 +t s^3 -4 s u t^2}{s t N}
\rq \absq{\bst} 
\nonumber \\
&& \hspace{-2.0cm}
+\lq 
\frac{(u-t) t (4t^2-5su) N}{s^2}
+\frac{t (2 t^2 +3 t s +5 s^2)}{s N}
\rq 
\bsu^\dag \bst 
\nonumber \\
&& \hspace{-2.0cm}
+\lq
\frac{2(u-t)(2 t^2 -su) N}{t}
+\frac{2 s (t^2+s^2)}{t N}
\rq 
\btu^\dag \bst 
\nonumber \\
&& \hspace{-2.0cm}
+ \lq
(6 t+s)N 
+\frac{s (t+3 s)}{ t N} 
\rq  
\btu^\dag \fbub{t,s} 
\nonumber \\
&& \hspace{-2.0cm}
+ \lq
\frac{-(6 u^2 + s^2) N}{t} +
\frac{2 (5 s^2-8tu )}{N t}
\rq 
\ep \btu^\dag \bubs 
\nonumber \\
&& \hspace{-2.0cm}
+ \lq
\frac{(12 t^3 +17 t^2 s +13 t s^2 +2 s^3) N}{ 2 s^2}
+\frac{(t+ 3 s)(2 t^2 + t s +2 s^2)}{ 2 s t N}
\rq \bst^\dag \fbub{t,s}  
\nonumber \\
&& \hspace{-2.0cm}
+ \lq
\frac{(12 t^2 + 11 t s + 9 s^2)t N}{ 2 s^2}
+\frac{(t+ 3 s) t}{s N}
\rq \bsu^\dag \fbub{t,s}
\nonumber \\
&& \hspace{-2.0cm}
+ \lq
-\frac{(12 t^2 +33 t s + 23 s^2) t N}{2 s^2}
+\frac{2 ( 8 t^2+14 t s+ 11 s^2) t}{s^2 N}
\rq \ep \bsu^\dag \bubs
\nonumber \\
&& \hspace{-2.0cm}
+\lq 
-\frac{(12 t^4 + 39 t^3 s + 38 t^2 s^2 + 31 t s^3 + 14 s^4) N}{2 s^2 t}
\right.
\nonumber\\
&& \hspace{-2.0cm}
+\left.\frac{16 t^4 + 36 t^3 s + 34 t^2 s^2 +21 t s^3 +10 s^4}{s^2 t  N}
\rq  \ep \bst^\dag \bubs,
\end{eqnarray}
\begin{eqnarray}
\MAB(s, t, u) &=& 
2 \bnot^2 
\left\{
\lq  \frac{t^2+u^2}{s^2}\rq \fbub{t,s}^\dag \fbub{u,s} -\frac{2 u t}{s^2} \ep^2 
\absq{\bubs} 
\right\} \nonumber \\
&& \hspace{-2.0cm} 
+\bnot \left\{
 \lq \frac{2 (t^2+8ut+u^2) N}{s^2} +\frac{2}{N}
\rq \ep^2 \absq{\bubs} \right.
\nonumber \\
&& \hspace{-2.0cm}
+\lq 
-\frac{(6 t^2+s^2) N}{s^2} + \frac{2 (5 s^2-8 t u)}{s^2 N} 
\rq 2\,  \ep  \Re \left[ \bubs^\dag \fbub{t,s} \right]
\nonumber \\
&& \hspace{-2.0cm}
+\lq
\frac{2 (u-t) (2 t^2-s u) N}{s^2}
+ \frac{2 (t^2+s^2)}{s N}  
\rq  \, 2 \Re\left[
\bst^\dag \fbub{u,s}
\right]
\nonumber \\
&& \hspace{-2.0cm}
+ \left.\lq
\frac{(5s^2-12tu)N}{s^2} +\frac{5}{N}
\rq \fbub{t,s}^\dag \fbub{u,s} 
\right\}
\nonumber \\
&& \hspace{-2.0cm}
+\lq
-\frac{4 (s^2-ut) (t-u)^2 N^2}{s^2}
+\frac{4 s^2}{N^2}
\rq \bst^\dag \bsu
\nonumber \\
&& \hspace{-2.0cm}
+\lq
\frac{3(s^2-3tu)N^2}{s^2}+\frac{15}{2}   
\rq \fbub{t,s}^\dag \fbub{u,s}
\nonumber \\
&& \hspace{-2.0cm}
+\lq
\frac{(u-t)(6t^2+8ts+5s^2) N^2}{s^2}
+\frac{5t^2+2ts+3s^2}{s}
+\frac{2s-t}{N^2}
\rq \, 2\Re\left\{ 
\bst^\dag \fbub{u,s}
\right\} 
\nonumber \\
&& \hspace{-2.0cm}
+\lq
\frac{(t-u)(6 t^2 -3 t s +s^2)N^2}{s^2}
-\frac{32 t^3+35 t^2 s +22 t s^2+11s^3}{s^2}
\right.
\nonumber \\
&& \hspace{-2.0cm}
\left.
+\frac{2 (4 t^2 +5 s^2)}{s N^2}
\rq 2\, \ep \Re \left\{ \bst^\dag \bubs \right\}
\nonumber \\
&& \hspace{-2.0cm}
+\lq
\frac{3 (2s^2+5ts-6t^2 ) N^2}{2 s^2}
+\frac{48t^2+5ts -3s^2}{2 s^2}
+\frac{4 (t+3 s)}{s N^2}
\rq 2 \, \ep \Re \left\{\fbub{t,s}^\dag  \bubs \right\} 
\nonumber \\
&& \hspace{-2.0cm}
+\lq
\frac{(7s^2-36tu)N^2}{2s^2}
-\frac{43s^2-48tu}{s^2}
+\frac{47s^2-64tu}{s^2N^2} 
\rq\, \ep^2 \absq{\bubs}
\end{eqnarray}
and
\begin{eqnarray}
\label{eq:FAA}
\MAA(s, t, u) &=& \bnot^2 
\left\{
\frac{4ut}{s^2} \ep^2 \absq{\bubs} +
\frac{2u(t^2+u^2)}{s^2 t} \absq{\fbub{t,s}}
\right\}
\nonumber \\
&& \hspace{-2.0cm}
+\bnot \left\{
\lq 
\frac{2u(6t+s)N}{s^2}
+\frac{2u(t+3s)}{stN}
\rq \absq{\fbub{t,s}} \right.
\nonumber \\
&& \hspace{-2.0cm}
+\lq
-\frac{u(6t^2+12ts+7s^2)N}{s^2t}
+\frac{2u(5s^2-8tu)}{s^2tN}
\rq 2\,\ep   \Re \left[ \bubs^\dag \fbub{t,s}\right]
\nonumber \\
&& \hspace{-2.0cm}
+\lq
\frac{2u(u-t)(2t^2-su)N}{s^2t}
+\frac{2u(t^2+s^2)}{stN}
\rq \, 2\Re \left[ \bst^\dag \fbub{t,s}\right]
\nonumber \\
&& \hspace{-2.0cm}
+\left.\lq
\frac{4u(s-3t)N}{s^2}
+\frac{4u}{sN}
\rq \ep^2 \absq{\bubs}
\right\}
\nonumber \\
&& \hspace{-2.0cm}
+\lq
\frac{(t-u)(6t^4+21t^3s+25t^2s^2+19ts^3+7s^4)N^2}{s^2tu} \right.
\nonumber \\
&& \hspace{-2.0cm}
-\frac{32t^5+99t^4s+130t^3s^2+100t^2s^3+58ts^4+17s^5}{s^2tu} 
\nonumber \\
&& \hspace{-2.0cm}
\left.
+\frac{2(4t^4-8t^2su+8ts^3+5s^4)}{stuN^2}
\rq 2\,\ep \Re \left\{ \bst^\dag \bubs \right\}
\nonumber \\
&& \hspace{-2.0cm}
+\lq
\frac{2(t-u)^2(2t^4-4t^2su+2ts^3+s^4)N^2}{s^2tu}
+\frac{4(u-t)(t^4-2t^2su-s^3u)}{stu} \right.
\nonumber \\
&& \hspace{-2.0cm}
\left.
+\frac{2(s^4+t^2u^2)}{tuN^2}
\rq \absq{\bst}
\nonumber \\
&& \hspace{-2.0cm}
+\lq
\frac{(u-t)(6t^3+10t^2s+7ts^2+s^3)N^2}{s^2u}
+\frac{t^4-3t^3s-6t^2s^2-9ts^3-3s^4}{stu} \right.
\nonumber \\
&& \hspace{-2.0cm}
\left.
+\frac{s^2(t+3s)}{tuN^2}
\rq \, 2\Re \left\{\bst^\dag \fbub{t,s} \right\}
\nonumber \\
&& \hspace{-2.0cm}
+\lq
\frac{(36t^4+114t^3s+191t^2s^2+160ts^3+51s^4)N^2}{2s^2tu} \right.
\nonumber \\
&& \hspace{-2.0cm}
-\frac{2(24t^4+99t^3s+156t^2s^2+117ts^3+34s^4)}{s^2tu}
\nonumber \\
&& \hspace{-2.0cm}
\left.
+\frac{64t^4+192t^3s+241t^2s^2+162ts^3+51s^4}{s^2tuN^2}
\rq \ep^2 \absq{\bubs}
\nonumber \\
&& \hspace{-2.0cm}
+\lq
-\frac{(18t^3+57t^2s+42ts^2+7s^3)N^2}{2s^2u}
+\frac{48t^4+101t^3s+49t^2s^2-33ts^3-21s^4}{2s^2tu}
\right.
\nonumber \\
&& \hspace{-2.0cm}
\left.
+\frac{(t+3s)(4u^2+s^2)}{stuN^2} 
\rq 2 \, \ep  \Re \left\{ \bubs^\dag \fbub{t,s} \right\}
\nonumber \\
&& \hspace{-2.0cm}
+\lq
\frac{t(9t^2+12ts+5s^2)N^2}{s^2u}
+\frac{(t+3s)(s+3t)}{su}
+\frac{(t+3s)^2}{2tuN^2}
\rq \absq{\fbub{t,s}},
\end{eqnarray}
with
\begin{equation}
\fbub{t,s}=\bubt-(1+\ep) \bubs.
\end{equation}
Only the first term in the expansions of the box integrals are
required in Eqs.~(\ref{eq:FCC})--(\ref{eq:FAA}), where  we have systematically
discarded contributions of $\O{\ep}$.  The bubble integrals must be
expanded through to $\O{\ep^0}$.

In this section,  we have 
isolated the infrared  divergences of the one-loop amplitude into
the terms $\PA$ and $\PC$. Both functions depend on one-loop 
bubble integrals and  diverge as  $1/\ep^2$. The separation 
of the singular terms becomes  evident when we  complete  our basis
of one-loop master integrals with the finite box in $D=6-2\ep$. In the context
of one-loop integrals, this form for the divergences arises naturally.
The singular behaviour of the one-loop amplitude can 
also be predicted applying the formalism of~\cite{catani} which yields
\begin{equation}
\PAC(s, t, u) = -\frac{1}{st} \AA(\ep, s, u, t)
\end{equation}
and
\begin{equation}
\PCC(s, t, u) = -\frac{1}{su} \BB(\ep, s, t, u)-\frac{1}{st} \BB(\ep, s, u, t),
\end{equation}
where $\AA(\ep, s, t, u)$ and $\BB(\ep, s, t, u)$ are defined in
Eqs.~(\ref{eq:Adef}) and~(\ref{eq:Bdef}).  
The singular structure of $\PAC$ and $\PCC$ 
around $\ep=0$ precisely matches that given in Eq.~(\ref{eq:pa}) 
and~(\ref{eq:pc}) and the two formalisms differ only in the finite remainder.
In order to make the agreement explicit 
one could replace $\PA$($\PC$) with $\PAC$($\PCC$) in Eqs.~(\ref{eq:PSQUARE})
and~(\ref{eq:PLOOP}) with appropriate modifications due to the 
finite differences $\PA -\PAC$ and $\PC - \PCC$ 
in Eqs.~(\ref{eq:PLOOP}) and~(\ref{eq:finiteone}).

\section{Summary}
\label{sec:conc}
In this paper we presented the $\O{\as^4}$ QCD corrections to the $2 \to 2$
scattering processes $q \qb \to gg$, $gg \to \qb q$ and the associated
crossed processes $q g \to q g$ and $\qb g \to \qb g$ in the high energy
limit, where the quark masses can be ignored.  We computed renormalised
analytic expressions for the interference of the tree-level diagrams with the
two-loop ones and for the self-interference of one-loop graphs in the \MSbar\
scheme.  Throughout we employed conventional dimensional regularisation.

The renormalised matrix elements are infrared divergent and contain poles down
to $\O{1/\ep^4}$.    The singularity structure of one- and two-loop diagrams
has been thoroughly studied by Catani~\cite{catani} who provided a procedure
for predicting the  infrared behaviour of renormalised amplitudes.  The
anticipated pole structure  agrees exactly with that obtained by direct Feynman
diagram evaluation.  In fact Catani's method does not determine the 
$1/\ep$ poles
exactly, but expects that the remaining unpredicted
$1/\ep$ poles are  non-logarithmic and
proportional to  constants (colour factors, $\pi^2$ and $\zeta_3$).  We find
that this is indeed 
the case, and  the constant $H^{(2)}$ is given in Eq.~(\ref{eq:Htwo}).
This provides a very strong check on the reliability of our results.
Similarly, the infrared divergent structure of the squared one-loop diagrams 
we found by direct evaluation agrees with the expected pole structure. 

The pole structure of the two-loop contribution is given in 
Eq.~(\ref{eq:poles}) while expressions for the finite parts in the $s$-channel
and $u$-channels according to the colour decomposition of Eq.~(\ref{eq:zi})
are given in Secs.~\ref{subsec:uttex} and~\ref{subsec:sttex} respectively.
Similarly, the infrared divergent one-loop contributions along with the
remaining finite parts are detailed in Sec.~\ref{sec:one}.   The expressions
for the  two-loop pole structure and the singular and finite parts of the
self-interference of one-loop graphs are analytic and are given in terms of the
one-loop bubble graph and one-loop box graph in $D=6-2\ep$ dimensions. To evaluate
these 
formulae  in the appropriate physical region requires the insertion of the
series expansion of the one-loop graphs around $\epsilon = 0$.   These
expansions are given in Appendix~\ref{app:master_int}.

In this paper, we have concentrated only on QCD processes.  However, the QED
processes $e^+e^- \to \gamma \gamma$ and Compton scattering $e^-\gamma \to
e^-\gamma$ as well as $\gamma\gamma \to e^-e^+$ are also of interest.   
We note that the expressions given here are related to those for the
corresponding QED processes in the limit where the electron mass can be ignored
by the alteration of various colour factors.  We expect to address this in a
separate article.
  
The results presented here, together with those previously computed for
quark-quark scattering~\cite{qqQQ,qqqq,1loopsquare} are necessary ingredients
for 
the  next-to-next-to-leading order predictions for jet cross sections in
hadron-hadron collisions.  On their own, they are insufficient to make physical
predictions and much work remains to be done.    First, at the matrix element
level, similar expressions to those presented here  for gluon-gluon scattering
are needed.    Given the recent progress in the field, we anticipate that this 
problem will soon be solved.
Second, a systematic procedure for analytically canceling the
infrared divergences between the tree-level $2 \to 4$, the one-loop $2 \to 3$
and the $2\to 2$ processes  needs to be established for semi-inclusive jet
cross sections.  Again,
recent progress in determining the singular limits of tree-level matrix
elements when two particles are unresolved~\cite{tc,ds} and the soft and
collinear limits of one-loop amplitudes~\cite{sone,cone}, together with the
analytic cancellation of the infrared singularities in the somewhat simpler
case of $e^+e^- \to {\rm photon} + {\rm jet}$ at next-to-leading order
\cite{aude}, suggest that the technical problems will soon be solved for
generic $2 \to 2$ scattering processes.  There are additional problems due to
initial state radiation.  However, the recent steps taken towards the
determination of the three-loop splitting functions~\cite{moms1,moms2,Gra1} are
also promising.

Third,  a numerical implementation of the various contributions must be
developed.   The next-to-leading order programs for  three jet production that
have already been written provide a first step in this direction
\cite{trocsanyi,kilgore}. We are confident that the problem of the numerical
cancellation of residual infrared divergences will soon be addressed thereby
enabling the construction of numerical programs to provide
next-to-next-to-leading order QCD estimates of jet production in hadron
collisions.

\section*{Acknowledgements}

C.A. acknowledges the financial support of the Greek Government and
M.E.T. acknowledges financial support from CONACyT and the CVCP. 
We gratefully acknowledge the support of
the British Council and German Academic Exchange Service under ARC project
1050.  This work was supported in part by the EU Fourth Framework Programme
`Training and Mobility of Researchers', Network `Quantum Chromodynamics and
the Deep Structure of Elementary Particles', contract FMRX-CT98-0194
(DG-12-MIHT), in part by the U.S.~Department of Energy
under
Grant No.~DE-FG02-95ER40896 and in part by the University of Wisconsin
Research Committee with funds granted by the Wisconsin Alumni Research
Foundation.

\appendix
\section{One-loop master integrals}
\label{app:master_int}
In this appendix, we list the expansions for the one-loop box integrals in
$D=6-2\ep$.
We remain in the physical region $s>0$, $u,t < 0$, 
and write coefficients in terms of logarithms and polylogarithms that are
real in this domain.  More precisely, we use the notation of
Eqs.~(\ref{eq:xydef}) and~(\ref{eq:xydef1}) to define the arguments of the
logarithms and 
polylogarithms. The polylogarithms are defined as in
Eq.~(\ref{eq:lidef}).

We find that the box integrals have the expansion,
\begin{eqnarray}
\Bfin &=& \frac{ e^{\ep\gamma}
\Gamma  \left(  1+\epsilon \right)  \Gamma  
\left( 1-\epsilon \right) ^2 
 }{ 2s\Gamma  \left( 1-2 \epsilon  \right)   \left( 1-2 \epsilon  \right) } 
 \left(\frac{\mu^2}{s} \right)^{\ep}
  \Biggl\{
 \frac{1}{2}\lq\(\lnx-\lny\)^2+\pi^2 \rq\nonumber \\
&& 
 +2\ep \lq
 \Licx-\lnx\Libx-\frac{1}{3}\lnx^3-\frac{\pi^2}{2}\lnx \rq
\nonumber \\
&& 
-2\ep^2\Bigg[
\Lidx+\lny\Licx-\frac{1}{2}\lnx^2\Libx-\frac{1}{8}\lnx^4-\frac{1}{6}\lnx^3\lny+\frac{1}{4}\lnx^2\lny^2\nonumber
\\
&&\qquad
\qquad-\frac{\pi^2}{4}\lnx^2-\frac{\pi^2}{3}\lnx\lny-\frac{\pi^4}{45}\Bigg]
+ ( u \leftrightarrow t) \Biggr\} + \O{\ep^3},
\end{eqnarray}
and
\begin{eqnarray}
\label{eq:boxst}
{\rm Box}^6(s, t)&=&\frac{e^{\ep\gamma}\Gamma(1+\ep) \Gamma(1-\ep)^2}
{2 u\Gamma(1-2\ep)(1-2\ep)}\,\fu  \Biggl\{ \left(\Lx^2 +2 i\pi
\Lx\right)\nonumber \\
&&+\ep \Biggl[
\left(-2\Licx+2 \Lx \Libx  -\frac{2}{3} \Lx^3+2 \Ly \Lx^2-\pi^2 \Lx+2 \zeta_3
\right)\nonumber \\
&& \qquad \qquad +i\pi\left(2 \Libx +4 \Ly \Lx-\Lx^2-\frac{\pi^2}{3}\right)
\Biggr]\nonumber \\
&&+\ep^2 \Bigg[
\Biggl(2\Lidz+2\Lidy-2\Ly \Licx-2\Lx \Licy+(2\Lx\Ly-X^2-\pi^2)\Libx
\nonumber \\
&&\qquad+\frac{1}{3}\Lx^4-\frac{5}{3}\Lx^3\Ly+\frac{3}{2}\Lx^2\Ly^2+\frac{2}{3}\pi^2\Lx^2-2\pi^2\Lx\Ly+2\Ly\zeta_3+\frac{1}{6}\pi^4\Biggr)
\nonumber \\
&&\qquad \qquad + i\pi \biggl(
-2\Licx-2\Licy+2\Ly\Libx+\frac{1}{3}\Lx^3-2\Lx^2\Ly+3\Lx\Ly^2\nonumber \\
&& \qquad \qquad \qquad \qquad -\frac{\pi^2}{3}\Ly+2\zeta_3
\biggr)\Bigg] \Biggr\} + \O{\ep^3}.
\end{eqnarray}
${\rm Box}^6(s,u)$ is obtained from Eq.~(\ref{eq:boxst}) by exchanging $u$ and
$t$.

Finally, the one-loop bubble integral in $D=4-2 \epsilon$ dimensions 
is given by
\begin{equation} 
 \Bubl =\frac{ e^{\ep\gamma}\Gamma  \left(  1+\epsilon \right)  \Gamma  \left( 1-\epsilon \right) ^2 
 }{ \Gamma  \left( 2-2 \epsilon  \right)  \epsilon   } \fs.
\end{equation}

\end{document}